\documentclass[twocolumn,nofootinbib]{aastex62}
\usepackage{calc}
\usepackage{mathtools,graphicx}
\usepackage{pifont}
\usepackage{bm}
\usepackage{microtype}
\usepackage{booktabs}
\usepackage{times}
\usepackage[varg]{txfonts}
\usepackage[utf8]{inputenc}
\usepackage[caption=false]{subfig}
\usepackage{paralist}
\usepackage[normalem]{ulem}
\usepackage{multirow}
\usepackage{etoolbox}
\usepackage{xstring}
\usepackage{xparse}
\usepackage[nolist,nohyperlinks,printonlyused]{acronym}
\usepackage{dcolumn}

\definecolor{LinkColor}{rgb}{0.75, 0, 0}
\definecolor{CiteColor}{rgb}{0, 0.5, 0.5}
\definecolor{UrlColor}{rgb}{0, 0, 0.75}
\hypersetup{linkcolor=LinkColor}
\hypersetup{citecolor=CiteColor}
\hypersetup{urlcolor=UrlColor}
\allowdisplaybreaks
\DeclareFontFamily{OT1}{pzc}{}
\DeclareFontShape{OT1}{pzc}{m}{it}{<-> s * [1.10] pzcmi7t}{}
\DeclareMathAlphabet{\mathpzc}{OT1}{pzc}{m}{it}

\setcitestyle{notesep={}}

\AtBeginDocument{
    \newwrite\bibnotes
    \def\bibnotesext{Notes.bib}
    \immediate\openout\bibnotes=\jobname\bibnotesext
    \immediate\write\bibnotes{@CONTROL{REVTEX41Control}}
    \immediate\write\bibnotes{@CONTROL{%
    apsrev41Control,author="08",editor="1",pages="0",title="0",year="1"}}
     \if@filesw
     \immediate\write\@auxout{\string\citation{apsrev41Control}}
    \fi
}

\newtoggle{fullauthorlist}
\toggletrue{fullauthorlist}

\newtoggle{endauthorlist}
\toggletrue{endauthorlist}

\graphicspath{{./fig/}}

\usepackage{scrextend}
\usepackage{textgreek}
\usepackage{xspace}

\usepackage{graphicx}
\usepackage{makerobust}
\makeatletter
\@ifundefined{caption@xref}{}{%
  \MakeRobustCommand\caption@xref
}
\makeatother

\def\mnras{\ref@jnl{MNRAS}}             

\renewcommand{\today}{\number\day\space\ifcase\month\or
  January\or February\or March\or April\or May\or June\or
  July\or August\or September\or October\or November\or December\fi
  \space\number\year}

\newtoggle{cleancomments}
\togglefalse{cleancomments}

\newcommand{\NUMEVENTS}{39}

\newcommand{\IMRD}{\textsc{IMRPhenomD}\xspace}
\newcommand{\IMRP}{\textsc{IMRPhenomPv2}\xspace}
\newcommand{\IMRPHM}{\textsc{IMRPhenomPv3HM}\xspace}
\newcommand{\IMRPNRT}{\textsc{IMRPhenomPv2\_NRTidal}\xspace}
\newcommand{\IMRXPHM}{\textsc{IMRPhenomXPHM}\xspace}
\newcommand{\NRSur}{\textsc{NRSur7dq4}\xspace}

\newcommand{\linf}{\textsc{LALInference}\xspace}

\newcommand{\linfnest}{\textsc{LALInferenceNest}\xspace}
\newcommand{\lal}{\textsc{LALSuite}\xspace}
\newcommand{\gstlal}{\textsc{GstLAL}\xspace}
\newcommand{\pycbc}{\textsc{PyCBC}\xspace}
\newcommand{\cwb}{\textsc{cWB}\xspace}
\newcommand{\hanabi}{\textsc{hanabi}\xspace}
\newcommand{\bilby}{\textsc{Bilby}\xspace}
\newcommand{\bayestar}{\textsc{Bayestar}\xspace}

\newcommand{\gw}[1][]{gravitational wave#1 (GW#1)\renewcommand{\gw}[1][]{GW##1\xspace}\xspace}
\newcommand{\cbc}[1][]{compact binary coalescence#1 (CBC#1)\renewcommand{\cbc}[1][]{CBC##1\xspace}\xspace}
\newcommand{\bns}[1][]{binary neutron star#1 (BNS#1)\renewcommand{\bns}[1][]{BNS##1\xspace}\xspace}
\newcommand{\bbh}[1][]{binary black hole#1 (BBH#1)\renewcommand{\bbh}[1][]{BBH##1\xspace}\xspace}
\newcommand{\nsbh}[1][]{neutron star--black hole#1 (NSBH#1)\renewcommand{\nsbh}[1][]{NSBH##1\xspace}\xspace}
\newcommand{\pe}{parameter estimation (PE)\renewcommand{\pe}{PE\xspace}\xspace}
\newcommand{\pisn}{pair instability supernova (PISN)\renewcommand{\pisn}{PISN\xspace}\xspace}
\newcommand{\kde}[1][]{kernel density estimator#1 (KDE#1)\renewcommand{\kde}[1][]{KDE##1\xspace}\xspace}
\newcommand{\fap}[1][]{false-alarm probability#1 (FAP#1)\renewcommand{\fap}[1][]{FAP##1\xspace}\xspace}
\newcommand{\far}[1][]{false-alarm rate#1 (FAR#1)\renewcommand{\far}[1][]{FAR##1\xspace}\xspace}
\newcommand{\sgwb}{stochastic GW background (SGWB)\renewcommand{\sgwb}{SGWB\xspace}\xspace}
\newcommand{\snr}[1][]{signal-to-noise ratio#1 (SNR#1)\renewcommand{\snr}[1][]{SNR##1\xspace}\xspace}

\newcommand{\Msun}{\ensuremath{\text{M}_\odot}}
\newcommand{\Mc}{\ensuremath{\mathcal{M}}}

\newcommand{\posstatistic}{\mathcal{B}^{\rm overlap}}
\newcommand{\lensedhyp}{\mathcal{H}_{\rm SL}}
\newcommand{\notlensedhyp}{\mathcal{H}_{\rm U}}
\newcommand{\cohratio}{\mathcal{C}^\mathrm{L}_\mathrm{U}}
\newcommand{\Blu}{\mathcal{B}^\mathrm{L}_\mathrm{U}}

\newcommand{\BMLU}{\mathcal{B}_\mathrm{U}^\mathrm{ML}}
\newcommand{\zL}{z_\mathrm{l}}
\newcommand{\MzL}{M^z_\mathrm{L}}

\newcommand{\RateAboveTwo}{\ensuremath{4.9_{-1.3}^{+1.7}\times{10^{-4}}}}
\newcommand{\RateAboveThirty}{\ensuremath{3.5_{-0.4}^{+0.6}\times{10^{-5}}}}

\newcommand{\RateModelAGalaxyDouble}{\ensuremath{0.9\text{--}4.4\times{10^{-4}}}}
\newcommand{\RateModelAGalaxySingle}{\ensuremath{2.9\text{--}9.5\times{10^{-4}}}}
\newcommand{\RateModelAClusterDouble}{\ensuremath{0.4\text{--}1.8\times{10^{-4}}}}
\newcommand{\RateModelAClusterSingle}{\ensuremath{1.4\text{--}4.1\times{10^{-4}}}}

\newcommand{\RateModelBGalaxyDouble}{\ensuremath{1.0\text{--}23.5\times{10^{-4}}}}
\newcommand{\RateModelBGalaxySingle}{\ensuremath{2.5\text{--}45.2\times{10^{-4}}}}
\newcommand{\RateModelBClusterDouble}{\ensuremath{0.7\text{--}10.9\times{10^{-4}}}}
\newcommand{\RateModelBClusterSingle}{\ensuremath{1.6\text{--}19.9\times{10^{-4}}}}

\begin{document}

\title{Search for lensing signatures in the gravitational-wave observations from the first half of LIGO-Virgo's third observing run}

\iftoggle{endauthorlist}{
 \let\mymaketitle\maketitle
 \let\myauthor\author
 \let\myaffiliation\affiliation
 \author{The LIGO Scientific Collaboration and the Virgo Collaboration}
}{
 \iftoggle{fullauthorlist}{
  \author{R.~Abbott}
\affiliation{LIGO Laboratory, California Institute of Technology, Pasadena, CA 91125, USA}
\author{T.~D.~Abbott}
\affiliation{Louisiana State University, Baton Rouge, LA 70803, USA}
\author{S.~Abraham}
\affiliation{Inter-University Centre for Astronomy and Astrophysics, Pune 411007, India}
\author{F.~Acernese}
\affiliation{Dipartimento di Farmacia, Universit\`a di Salerno, I-84084 Fisciano, Salerno, Italy  }
\affiliation{INFN, Sezione di Napoli, Complesso Universitario di Monte S.Angelo, I-80126 Napoli, Italy  }
\author{K.~Ackley}
\affiliation{OzGrav, School of Physics \& Astronomy, Monash University, Clayton 3800, Victoria, Australia}
\author{A.~Adams}
\affiliation{Christopher Newport University, Newport News, VA 23606, USA}
\author{C.~Adams}
\affiliation{LIGO Livingston Observatory, Livingston, LA 70754, USA}
\author{R.~X.~Adhikari}
\affiliation{LIGO Laboratory, California Institute of Technology, Pasadena, CA 91125, USA}
\author{V.~B.~Adya}
\affiliation{OzGrav, Australian National University, Canberra, Australian Capital Territory 0200, Australia}
\author{C.~Affeldt}
\affiliation{Max Planck Institute for Gravitational Physics (Albert Einstein Institute), D-30167 Hannover, Germany}
\affiliation{Leibniz Universit\"at Hannover, D-30167 Hannover, Germany}
% ...

\collaboration{The LIGO Scientific Collaboration and the Virgo Collaboration}
\noaffiliation

 }{
 \author{The LIGO Scientific Collaboration and the Virgo Collaboration}
}
}

\date{13 May 2021}

\begin{abstract}
\noindent
We search for signatures of gravitational lensing in the gravitational-wave signals from compact binary coalescences
detected by Advanced LIGO and Advanced Virgo during O3a,
the first half of their third observing run.
We study:
1) the expected rate of lensing at current detector sensitivity
and the implications of a non-observation of strong lensing or a stochastic gravitational-wave background on the merger-rate density at high redshift;
2) how the interpretation of individual high-mass events would change if they were found to be lensed;
3) the possibility of multiple images due to strong lensing by galaxies or galaxy clusters;
and 4) possible wave-optics effects due to point-mass microlenses.
Several pairs of signals in the multiple-image analysis show similar parameters and, in this sense, are nominally consistent with the strong lensing hypothesis.
However, taking into account population priors, selection effects, and the prior odds against lensing, 
these events do not provide sufficient evidence for lensing.
Overall, we find
no compelling evidence for lensing in the observed gravitational-wave signals
from any of these analyses.

\end{abstract}

\section{Introduction}
\label{sec:intro}
Gravitational lensing occurs when a massive object bends spacetime in a way that focuses light rays toward an observer \citep[see][ for a review]{Bartelmann:2010fz}.
Lensing observations are widespread in electromagnetic astrophysics
and have been used for, among other purposes, making a compelling case for dark matter~\citep{Clowe:2003tk,Markevitch:2003at}, discovering exoplanets~\citep{Bond:2004qd}, and uncovering massive objects and structures that are too faint to be detected directly~\citep{Coe:2012wj}.

Similarly to light, when \gw[s] 
travel near a galaxy or a galaxy cluster, their trajectories curve, resulting in gravitational lensing~\citep{Ohanian:1974ys,Thorne:1982cv,Deguchi:1986zz,Wang:1996as,Nakamura:1997sw,Takahashi:2003ix}.
For massive lenses, this changes the \gw amplitude without affecting the frequency evolution~\citep{Wang:1996as,Dai:2017huk,Ezquiaga:2020gdt}.
Strong lensing, in particular, can also produce multiple images observed at the \gw detectors as repeated events separated by a time delay of minutes to months for galaxies~\citep{Ng:2017yiu,Li:2018prc,Oguri:2018muv}, and up to years
for galaxy clusters~\citep{Smith:2017mqu,Smith:2018gle,Smith:2019dis,Robertson:2020mfh,Ryczanowski:2020mlt}.
The detection of such strongly lensed \gw[s] has been forecast within this decade~\citep{Ng:2017yiu,Li:2018prc,Oguri:2018muv},
at design sensitivity of Advanced LIGO and Advanced Virgo,
assuming that \bbh[s] trace the star-formation rate density.
In addition, if \gw[s] propagate near smaller lenses such as stars or compact objects, microlensing may induce observable beating patterns in the waveform~\citep{Deguchi:1986zz,Nakamura:1997sw,Takahashi:2003ix,Cao:2014oaa,Jung:2017flg,Lai:2018rto,Christian:2018vsi,Dai:2018enj,Diego:2019lcd,Diego:2019rzc,Pagano:2020rwj,Cheung:2020okf,Mishra:2021abc}.
Indeed, lensing can induce a plethora of effects on \gw[s].

If observed, \gw lensing could enable numerous scientific pursuits,
such as localization of merging black holes to sub-arcsecond precision~\citep{Hannuksela:2020xor},
precision cosmography studies~\citep{Sereno:2011ty,Liao:2017ioi,Cao:2019kgn,Li:2019rns,Hannuksela:2020xor},
precise tests of the speed of gravity~\citep{Baker:2016reh,Fan:2016swi},
tests of the \gw['s] polarization content~\citep{Goyal:2020bkm},
and detecting intermediate-mass or primordial black holes~\citep{Lai:2018rto,Diego:2019rzc,Oguri:2020ldf}.

Here we perform a comprehensive lensing analysis of data from the first half of the third LIGO--Virgo observing run,
called O3a for short,
focusing on \cbc signals.
We begin by outlining the expected rate of strongly lensed events.
Strong lensing is rare, but magnified signals
enable us to probe a larger comoving volume,
thus potentially giving us access to more sources~\citep{Dai:2016igl,Ng:2017yiu,Smith:2017mqu,Li:2018prc,Oguri:2018muv,Smith:2018gle,Smith:2019dis,Robertson:2020mfh,Ryczanowski:2020mlt}.
We forecast the lensed event rates using standard lens and black hole population models (Sec.~\ref{sec:statistics}).
These expected rates are subject to some astrophysical uncertainty but are vital to interpreting our search results in later sections.

The rate of lensing can also be inferred from the \sgwb~\citep{Buscicchio:2020cij,Mukherjee:2020tvr,Buscicchio:2020bdq}.
Thus, we use the non-observation of strong lensing and the stochastic background
to constrain the \bbh merger-rate density and the rate of lensing at high redshifts.

In addition, lensing magnification biases the inferred \gw luminosity distance and source mass measurements, which could
lead to observations of apparently high-mass (or low-mass, when de-magnified) binaries~\citep{Dai:2016igl,Broadhurst:2018saj,Oguri:2018muv,Hannuksela:2019kle,Broadhurst:2020moy}.
Therefore, we analyze several LIGO--Virgo detections with unusually high masses under the alternative interpretation that they are lensed signals from lower-mass sources which have been magnified (Sec.~\ref{sec:individualevents}).

We then move on to search for signatures of lensing-induced multiple images,
which should appear as repeated similar signals, magnified and with waveform differences determined by the image type~\citep{Dai:2017huk,Ezquiaga:2020gdt}, 
separated in time by minutes to months (or even years).
Consequently, if an event pair is strongly lensed, we expect to infer consistent parameters for both events~\citep{Haris:2018vmn,Hannuksela:2019kle}.

We search for these multiple images by first comparing the posterior overlap between pairs of events occurring during the O3a period as reported in~\cite{GWTC2} (Sec.~\ref{sec:posterioroverlap}).
After identifying a list of candidates from the posterior-overlap analysis, we follow these up with more computationally expensive but more accurate joint-\pe procedures (Sec.~\ref{sec:jointpe}).
Next, we perform a targeted search for previously undetected counterpart images of known events in Sec.~\ref{sec:subthreshold},
images that could have fallen below the threshold of previous wide-parameter space \cbc searches~\citep[as discussed in][]{Li:2019osa,McIsaac:2019use,Dai:2020tpj}.
Finally, we search for microlensing induced by point-mass lenses in the intermediate and low mass range, including wave-optics effects (Sec.~\ref{sec:microlensing}).

Several searches for \gw lensing signatures have already been performed in the data from the first two observing runs O1 and O2~\citep{Hannuksela:2019kle,Li:2019osa,McIsaac:2019use,Pang:2020qow,Liu:2020par,Dai:2020tpj}, including strong lensing and microlensing effects.
We will discuss these previous studies in the appropriate sections.
Given the growing interest in \gw lensing and the existing forecasts, 
an analysis of the most recent \gw observations for lensing effects is now timely.

Results of all analyses in this paper and associated data products can be found in~\citet{datarelease}.
\gw strain data and posterior samples for all events from GWTC-2
are available~\citep{gwosc:gwtc2} from the Gravitational Wave Open Science Center~\citep{Abbott:2019ebz}.

\section{Data and events considered}
\label{sec:events}
The analyses presented here use data taken during O3a by the Advanced LIGO~\citep{TheLIGOScientific:2014jea} and Advanced Virgo~\citep{TheVirgo:2014hva} detectors.  
O3a extended from 2019 April 1 to 2019 October 1.
Various instrumental upgrades have led to more sensitive data,
with median \bns inspiral ranges~\citep{PhysRevD.85.122006} increased by a factor of 1.64 in LIGO Hanford, 1.53 in LIGO Livingston, and 1.73 in Virgo
compared to O2~\citep{GWTC2}.
The duty factor for at least one detector being online was 97\%; for any two detectors being online at the same time it was 82\%; and for all three detectors together it was 45\%.
Further details regarding instrument performance and data quality for O3a are available in \citet{GWTC2,Buikema:2020dlj}.

The LIGO and Virgo detectors used a photon recoil based
calibration~\citep{Karki:2016pht,Cahillane:2017vkb,Viets:2017yvy}
resulting in a complex-valued, frequency-dependent detector
response with typical errors in magnitude of $7$\% and
$4$ degrees in phase~\citep{Sun_2020, Acernese2018} in the
calibrated O3a strain data.

Transient noise sources, referred to as glitches, contaminate the data and can affect the confidence of candidate detections.
Times affected by glitches are identified so that searches for \gw events can exclude (veto) these periods of poor data 
quality~\citep{TheLIGOScientific:2016zmo,LIGOScientific:2019hgc,Davis:2021ecd,nguyen2021environmental,Fiori2020}.
In addition, several known noise sources are subtracted from the data
using information from witness auxiliary sensors~\citep{Driggers:2018gii,Davis:2018yrz}.

Candidate events, including those reported in~\citet{GWTC2}
and the new candidates found by the searches for sub-threshold counterpart images in Sec.~\ref{sec:subthreshold} of this paper,
have undergone a validation process to evaluate if instrumental artifacts could affect the analysis;
this process is described in detail in Sec. 5.5 of~\cite{Davis:2021ecd}.
This process can also identify data quality issues that need further mitigation for individual events,
such as subtraction of glitches~\citep{Cornish:2014kda} and non-stationary noise couplings~\citep{Vajente:2019ycy},
before executing \pe algorithms.
See Table V of \citet{GWTC2} for the list of events requiring such mitigation.

The GWTC-2 catalog~\citep{GWTC2} contains \NUMEVENTS{} events from O3a
(in addition to the 11 previous events from O1 and O2)
with a \far below two per year,
with an expected rate of false alarms from detector noise less than 10\%~\citep{GWTC2}.
We neglect the potential contamination in this analysis.
These events were identified by three search pipelines:
one minimally modeled transient search \cwb~\citep[][]{Klimenko:2004qh,Klimenko:2005xv,Klimenko:2006rh,Klimenko:2011hz,Klimenko:2015ypf}
and the two matched-filter searches
\gstlal~\citep{Sachdev:2019vvd, Hanna:2019ezx, Messick:2016aqy}
and \pycbc~\citep{Allen:2005fk, Allen:2004gu, Canton:2014ena, Usman:2015kfa,Nitz:2017svb}.
Their parameters were estimated through Bayesian inference using the
\linf~\citep{Veitch:2014wba}
and \bilby~\citep{Ashton:2018jfp, Smith:2019ucc,Romero-Shaw:2020owr}
packages.
Both the matched-filter searches and \pe use a variety of
\cbc waveform models which generally combine knowledge from
post-Newtonian theory,
the effective-one-body formalism
and numerical relativity
\citep[for general introductions to these approaches, see][ and references therein]{Blanchet:2013haa,Damour:2016bks,Palenzuela:2020tga,Schmidt:2020ekt}.
The analyses in this paper rely on the same methods,
and the specific waveform models and analysis packages used
are described in each section.

Most of the \NUMEVENTS{} events from O3a are most probably \bbh[s],
while three
(GW190425, GW190426\_152155, and GW190814)
have component masses below 3\,$\Msun$~\citep{GW190425, GWTC2, GW190814},
thus potentially containing a neutron star.
We consider these \NUMEVENTS{} events in most of the analyses in this paper,
except in the magnification analysis (Sec.~\ref{sec:individualevents}),
which concerns only six of the more unusual events,
and the microlensing analysis (Sec.~\ref{sec:microlensing}),
which focuses on the 36 clear \bbh events only.

Specifically, we use the following input data sets for each analysis.
The magnification analysis in Sec.~\ref{sec:individualevents}
and
posterior-overlap analysis in Sec.~\ref{sec:posterioroverlap}
start from the Bayesian inference posterior samples released with GWTC-2~\citep{gwosc:gwtc2}.
The joint-\pe analyses in Sec.~\ref{sec:jointpe}
and microlensing analysis in Sec.~\ref{sec:microlensing}
reanalyze the strain data in short segments around the event times,
available from the same data release,
with data selection and noise mitigation choices matching those of the \pe analyses in~\citet{GWTC2}.
In addition, the searches for sub-threshold counterpart images in Sec.~\ref{sec:subthreshold}
cover the whole O3a strain data set,
using the same data quality veto choices as in~\citet{GWTC2}
but a strain data set consistent with the \pe analyses:
the final calibration version of LIGO data~\citep{Sun_2020}
with additional noise subtraction~\citep{Vajente:2019ycy}.

\section{Lensing statistics}
\label{sec:statistics}
In this section, we first forecast the number of detectable strongly lensed events (Sec.~\ref{sub:lensing_rate_strong}). 
Then, we infer upper limits on the rate of strongly lensed events using two different methods;
the first uses only the non-detection of resolvable strongly lensed BBH events (Sec.~\ref{ssec:implications_of_non_detection}),
while the second leverages additionally the non-observation of the \sgwb (Sec.~\ref{sub:lensing_rate_sgwb})~\citep{Callister:2020arv,Abbott:2021xxi}.
Since the background would originate from higher redshifts, this second method complements the first method.

Throughout this section, we model the mass distribution of \bbh[s] following the results for the \textsc{Power Law + Peak model} of~\cite{GWTC2:rates}.
We consider two distinct models of the \bbh merger rate density.
Model A brackets most of the population synthesis results~\citep{Eldridge:2018nop,Neijssel:2019irh,Boco:2019teq,Santoliquido:2020axb} corresponding to Population I and II stars  
while Model B assumes the Madau--Dickinson ansatz~\citep{Madau:2014bja} where the rate peaks at a particular redshift.
For consistency with previous analyses \citep[e.g.,][]{Abbott:2021xxi},
we take the Hubble constant from Planck 2015 observations to be $H_0=67.9\ {\rm km\ s^{-1}\ Mpc^{-1}}$ \citep{Planck_2015}.
Detailed discussion on both models and their respective parametrization is given in Appendix~\ref{app:lens_statistics_supplementary}. 
The obtained rates are subject to uncertainty because of their dependence on the merger rate density, which is model-dependent and only partially constrained.
They are nevertheless vital to interpreting our search results in later sections (see Sec.~\ref{sec:multiimages}).

\subsection{Strong lensing rate}
\label{sub:lensing_rate_strong}
We predict the rate of lensing using the standard methods outlined in the literature~\citep{Ng:2017yiu,Li:2018prc,Oguri:2018muv,Xu:2021bfn,Mukherjee:2021qam,Wierda:2021upe}, at galaxy and galaxy-cluster lens mass scales.
To model the lens population, we need to choose a density profile and a mass function. 
We adopt the singular isothermal sphere (SIS) density profile for both galaxies and clusters. 
Moreover, we use the velocity dispersion function (VDF) from the Sloan Digital Sky Survey \citep[SDSS;][]{Choi:2007a} for galaxies and the halo mass function from \citet{Tinker:2008ff} for clusters which have been used in other lensing studies as well \citep[e.g.,][]{Oguri:2010a,Robertson:2020mfh}.
The SIS profile can well describe galaxies. 
However, the mass distribution of clusters tends to be more complicated. 
Nevertheless, 
 \cite{Robertson:2020mfh} have demonstrated that
the SIS model can reproduce the lensing rate predictions from a study of numerically simulated cluster lenses.
Thus, we adopt the same model.  
Under the SIS model, we obtain two images with different magnifications
and arrival times. 
The rate of strong lensing is
\begin{equation}
\begin{split}
 \mathcal{R}_{\rm lens}= \int
& \frac{\mathrm{d}N(M_{\mathrm{h}},\zL)}{\mathrm{d}M_{\mathrm{h}}} \frac{\mathrm{d}D_{\mathrm{c}}} {\mathrm{d}\zL}
\frac{\mathcal{R}_{\mathrm{m}}(z_{\mathrm{m}})}{1+z_{\mathrm{m}}} \frac{\mathrm{d}V_{\mathrm{c}}}{\mathrm{d}z_{\mathrm{m}}}\, \sigma(M_{\mathrm{h}},\zL,z_{\mathrm{m}},\rho, \rho_{\mathrm{c}}) \\
&\times p(\rho|z_{\mathrm{m}}) ~\mathrm{d}\rho~\mathrm{d}z_{\mathrm{m}}~\mathrm{d}\zL~\mathrm{d}M_{\mathrm{h}} \,, \label{eq:rates1}
\end{split}
\end{equation}
where $\mathrm{d} N(M_\mathrm{h},\zL) / \mathrm{d}M_{\mathrm{h}}$ is the differential comoving number density of lensing halos in a halo mass shell at lens redshift $\zL$,
$D_{\mathrm{c}}$ and $V_\mathrm{c}$ are the comoving distance and volume, respectively, at a given redshift,
$\mathcal{R}_\mathrm{m}(z_\mathrm{m})$ is the total comoving merger rate density at redshift $z_\mathrm{m}$,
(1+$z_\mathrm{m}$) accounts for the cosmological time dilation,
$p(\rho\mid z_{\rm m})$ is the distribution of \snr at a given redshift, 
and $\sigma$ is the lensing cross-section (Appendix~\ref{app:lens_statistics_supplementary}).
Throughout this section and in Sec.~\ref{sub:lensing_rate_sgwb} we choose a network \snr threshold $\rho_\mathrm{c}=8$ as a point estimator of the detectability of \gw signals. 
We find it to be consistent with the search results in~\citet{GWTC2} and in Sec.~\ref{sec:subthreshold}, and we estimate its impact to be subdominant with respect to other source of uncertainties.

\begin{table*} 
\caption{
 \label{tab:lensedrates0_min}
 Expected fractional rates of observable lensed double events at current LIGO--Virgo sensitivity.
}
\begin{center}
\begin{tabular}{lcccc}
\hline\hline
 Merger Rate Density  & \multicolumn{2}{c} {Galaxies}  & \multicolumn{2}{c}  {Galaxy Clusters}\\ 
Model &  $R_{\rm D}$ & $R_{\rm S}$ & $R_{\rm D}$ & $R_{\rm S}$ \\ 
\hline\hline
 A       &  \RateModelAGalaxyDouble   &   \RateModelAGalaxySingle  & \RateModelAClusterDouble   &  \RateModelAClusterSingle \\
 B       & \RateModelBGalaxyDouble   &   \RateModelBGalaxySingle  &  \RateModelBClusterDouble   &  \RateModelBClusterSingle \\ 
\hline
\end{tabular}
\end{center}
\tablecomments{This table lists the relative rates of lensed double events expected to be observed by LIGO--Virgo at the
current sensitivity where both of the lensed events are detected ($R_{\rm D}$) and only one of the lensed events is detected ($R_{\rm S}$)
above the \snr threshold.
For Model A, the range corresponds to the bracketing function (see Appendix~\ref{app:lens_statistics_supplementary})
and for Model B, the rates encompass a 90~per~cent credible interval.
We show the rate of lensing by galaxies ($\sigma_{\rm vd}=10$--$300 \,{\rm km~s}^{-1}$)
and galaxy clusters ($\log_{10}(M_{\rm halo}/M_\odot) \sim 14$--$16$)
separately. 
Besides their usage for forecasts, the fraction of lensed events allows us to interpret the prior probability of the strong lensing hypothesis, which we require to identify lensed events confidently.
}
\end{table*}

In Table~\ref{tab:lensedrates0_min}, we show our estimates of the relative rate of lensing
assuming different models (Models A and B) for the merger rate density.
The results are shown separately for galaxy-scale (G) and cluster-scale (C) lenses.  
Furthermore, these rates are calculated for events that are doubly lensed and for two cases:
when only a single event (i.e., the brighter one) is detected (S),
and when both of the doubly lensed events are detected (D).
The expected fractional rate of lensing (lensed to unlensed rate), which will be necessary for the multi-image analyses (Sec.~\ref{sec:multiimages}), ranges from $\mathcal{O}(10^{-3}$--$10^{-4})$, depending on the merger rate density assumed.
We estimate the fractional rate of observed double (single) events for galaxy-scale lenses in the range 
 \RateModelAGalaxyDouble~(\RateModelAGalaxySingle) 
when using Model A for the merger rate density. 
Similarly, for cluster-scale lenses, the fractional rate is estimated to be in the range of 
\RateModelAClusterDouble~(\RateModelAClusterSingle) 
much rarer than the rates at galaxy scales. 
These estimates suggest that observing a lensed double image is unlikely at the current sensitivity of the LIGO--Virgo network of detectors. 
Nevertheless, at design sensitivity and with future upgrades, standard forecasts suggest that the possibility of observing such events might become significant~\citep{Ng:2017yiu,Li:2018prc,Oguri:2018muv,Xu:2021bfn,Mukherjee:2021qam,Wierda:2021upe}.
Our lensing rates are consistent with those predicted for
singular isothermal ellipsoid models \citep[e.g.,][]{Oguri:2018muv,Xu:2021bfn,Wierda:2021upe}.
The main uncertainty in the rate estimates derives from the uncertainties in the merger-rate density at high redshift. 

Depending on the specific distribution of lenses and the source population, the time delays between images can change.
Models favoring galaxy lensing produce minutes to perhaps months of time delay, while galaxy cluster lensing can produce time delays up to even years. 
However, the time delay distribution for galaxy cluster lenses is more difficult to model accurately, owing to the more complex lensing morphology.

Since the merger rate density at high redshift is observationally constrained only by the absence of the \sgwb, these rates are subject to uncertainty.
Nevertheless, standard theoretical models will still produce useful forecasts.
We will later refer to these rate estimates in the relevant sections (see Sec.~\ref{sec:multiimages}).

\subsection{Implications from the non-observation of strongly lensed events} 
\label{ssec:implications_of_non_detection}
Motivated by the absence of evidence for strong lensing (Sec.~\ref{sec:multiimages}), we assume that no strong lensing has occurred, in order to constrain the merger rate density at high redshift. 
We use the standard constraints on the merger rate density at low redshift from the LIGO--Virgo population studies~\citep{GWTC2:rates}. 
We assume the Madau--Dickinson form for the merger rate density (Model B). 
This model's free parameters include the local merger rate density, the merger rate density peak, and the power-law slope. 
\begin{figure}
    \centering
    \includegraphics[width=\columnwidth]{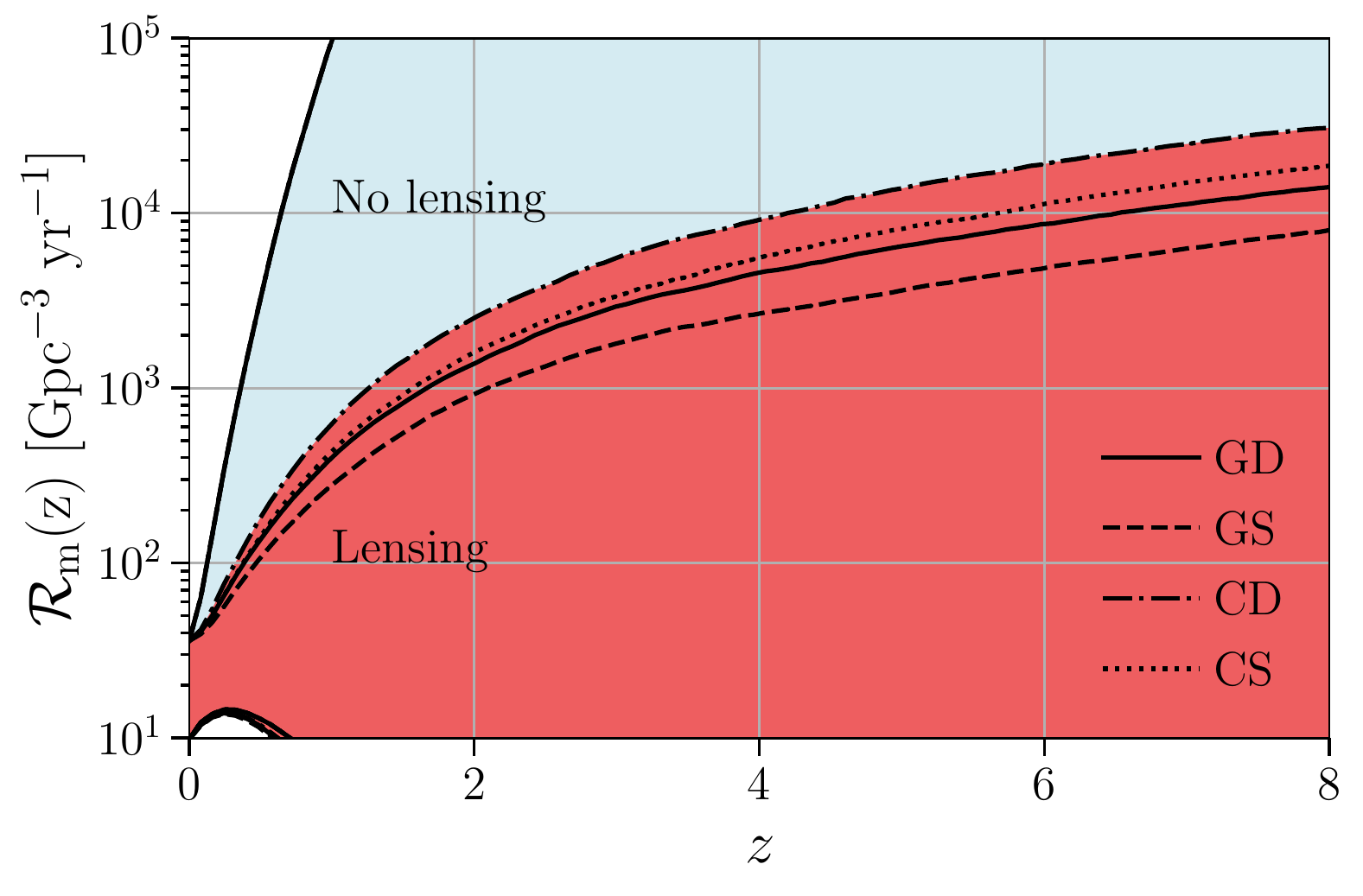}
    \caption{
	Merger rate density as a function of redshift based on the GWTC-2 results without lensing constraints (blue) and with lensing (red) included in LIGO--Virgo detections. 
        We show the results for galaxy-scale lenses (G) and cluster-scale lenses (C) separately. Furthermore, S (or D) correspond to doubly lensed events where single (or double) events are detected.  
	Because lensed detections occur at higher redshifts than unlensed events, their non-observation can be used to constrain mergers at higher redshifts.
    The results without lensing do not include constraints derived from the absence of a \sgwb.
	}
    \label{fig:merger_rate_constraints_from_strong_lensing}
\end{figure}
The non-observation of lensing constrains the merger-rate density at high redshift, which is unconstrained by the low-redshift observations alone (Fig.~\ref{fig:merger_rate_constraints_from_strong_lensing}).
These lensing constraints are complementary to the current strictest high-redshift limits obtained through \sgwb non-observation~\citep{Abbott:2021xxi}.

\subsection{Constraints from stochastic background}
\label{sub:lensing_rate_sgwb}

We can also constrain the redshift evolution of the merger rate density from the reported non-observation of the \sgwb from \bbh[s]~\citep{Callister:2020arv, Abbott:2021xxi}.
This, in turn, provides constraints on the relative abundance of distant mergers, which are more likely to undergo lensing. 
Thus, the non-observation of the \sgwb can inform the estimate of the probability of observing lensed \bbh mergers~\citep{Buscicchio:2020cij,Mukherjee:2020tvr}.

Following~\cite{Buscicchio:2020cij}, we forecast constraints on the merger rate density in O3 using up-to-date constraints on the mass distribution and redshift evolution of \bbh mergers obtained from the latest detections~\citep{LIGOScientific:2018jsj, GWTC1, GWTC2,GWTC2:rates},
as well as those inferred from current upper limits on the \sgwb, given its non-observation~\citep{Abbott:2021xxi}.

While the measured parameters for each merger (redshifts, source masses) are potentially biased by lensing, 
as discussed in Sec.~\ref{sec:individualevents}, 
we express all quantities as functions of non-biased merger redshift $z_\mathrm{m}$ and chirp mass $\mathcal{M}$~\citep{Buscicchio:2020cij} for consistency with other sections. 
However, following \citet{Buscicchio:2020cij}, we do not assume as prior information that lensing is not taking place.
Instead, we include the magnification bias self-consistently in the analysis, by imposing population constraints in apparent masses and redshifts.

\begin{figure}[t]
\includegraphics[width=1.0\columnwidth]{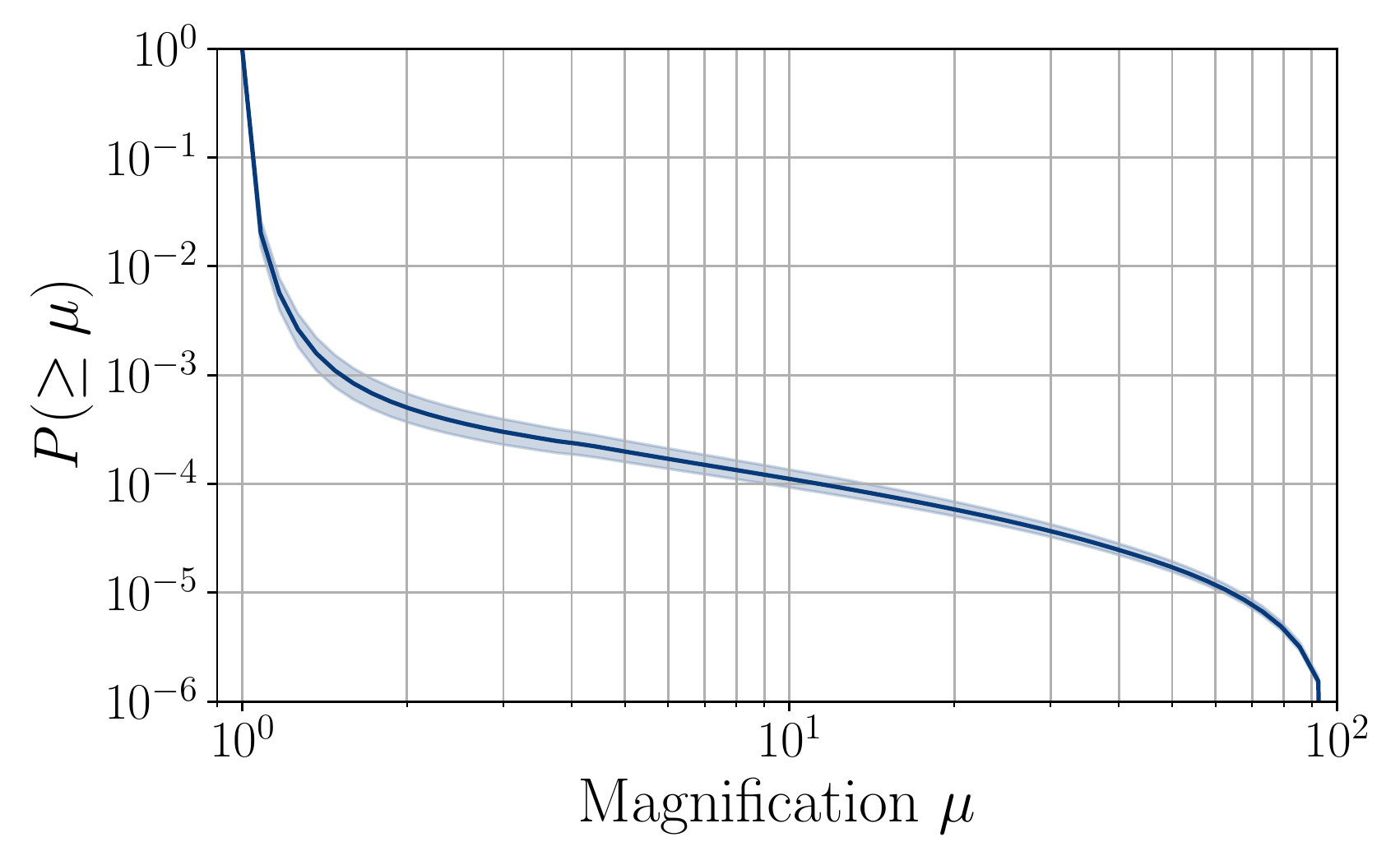}
\caption{\label{fig:cumulative}
Cumulative fraction of lensed detectable \bbh mergers at any redshift with magnification greater than $\mu$,
constrained by the non-observation of the \sgwb.
The solid line shows the value obtained from the median \bbh merger rate density posterior.
The shaded region corresponds to the $90\%$ credible interval. 
Fewer than $1$ in $10^3$ events are expected to be lensed with magnification $\mu>2$, on average.
Significantly higher magnifications (e.g., $\mu>30$) are suppressed by a further factor of $10$. 
The results here show the probability of \emph{observing} an event above a given magnification, which includes the merger-rate density and magnification bias information.
}
\end{figure}

We model the differential lensing probability following~\cite{Dai:2016igl}.
The differential merger rate in a redshift and magnification shell is
\begin{equation}\label{eq:diffmergerrate}
\begin{split}
\frac{\mathrm{d}^{2}\mathcal{R}}{\mathrm{d}z_\mathrm{m}\mathrm{d}\ln\mu}=&\frac{\mathrm{d}P(\mu\mid z_\mathrm{m})}{\mathrm{\mathrm{d}\ln\mu}} \frac{4\pi D_{\mathrm{c}}^{2}\left(z_\mathrm{m}\right)}{H_{0}\left(1+z_\mathrm{m}\right)E\left(z_\mathrm{m}\right)} \\
\times&\! \int\! \mathrm{d}{m}_{1}\mathrm{d}{m}_{2}\frac{\mathrm{d}^{3}{\mathcal{R}_\mathrm{m}(z_\mathrm{m})}}{\mathrm{d}{m}_{1}\mathrm{d}{m}_{2}\mathrm{d}{z_\mathrm{m}}}p\left(\rho\!>\!\rho_c\!\mid {m}_{1},{m}_{2},{z_\mathrm{m}},\mu\right)\!\,,
\end{split}
\end{equation}
where $\mathrm{d}^{3}{\mathcal{R}_\mathrm{m}(z_\mathrm{m})}/\mathrm{d} {m}_{1} \mathrm{d} {m}_{2} \mathrm{d} {z_\mathrm{m}}$ is the differential merger rate density; 
$p\left(\rho\!>\!\rho_\mathrm{c}\!\mid{m}_{1},{m}_{2},{z_\mathrm{m}}, \mu\right)$ provides the probability of observing mergers
with source masses $m_1,m_2$, redshift $z_\mathrm{m}$, and magnified by a factor $\mu$
above a fixed network SNR threshold $\rho_\mathrm{c}=8$, integrated over the population distribution of source parameters; 
the factor $4 \pi D_{\mathrm{c}}^{2}(z_\mathrm{m})/[H_{0}(1+z_\mathrm{m}) E(z_\mathrm{m})]$ gives the comoving volume of a redshift shell in an expanding Universe (taking into account the redshifted rate definition with respect to the source frame);
and $\mathrm{d} P(\mu \mid z_\mathrm{m})/(\mathrm{d} \ln \mu)$ is the lensing probability.
However, as noted by~\citet{Dai:2016igl}, the differential magnification probability
at $0.9<\mu<1.1$ and $z_m<2$ is affected by relative uncertainties up to $40\%$.
We therefore consider magnified detections only ($\mu>1$), which are subject to less uncertainty, 
and normalize our results accordingly.
We then integrate the differential merger rate (Eq.~\ref{eq:diffmergerrate}) over redshift and magnifications in $\left[\mu,\mu_{\max}\right]$ 
and divide it by the total rate of magnified detections.
By doing so, we obtain the cumulative fraction of detected lensed events at any redshift with magnifications larger than $\mu$.

We show the result in Fig.~\ref{fig:cumulative}. 
We find the observation of lensed events to be unlikely, with the fractional rate at $\mu>2$ being \RateAboveTwo.
More significantly magnified events are even more suppressed, with a rate of
\RateAboveThirty~at $\mu>30$.
These estimates suggest that most binary mergers that we observe are not strongly lensed. 
However, as projected in~\citet{Buscicchio:2020cij, Mukherjee:2020tvr}, at design sensitivity, the same probability will be enhanced, as a widened horizon will probe the merger rate density deeper in redshift.

Comparing the above predictions with the expected fractional rates $R_S$ of single lensed detections with Model B in Table.~\ref{tab:lensedrates0_min}, the predictions agree within a factor of 5 for the relative rate of lensing.
The differences are due to a different underlying lens model and partly to the inclusion of de-magnified events in Sec.~\ref{sub:lensing_rate_strong}.

\section{Analyzing high-mass events} 
\label{sec:individualevents}
If a \gw signal is strongly lensed, it will receive a magnification $\mu$ defined such that the \gw amplitude increases by a factor $|\mu|^{1/2}$ relative to an unlensed signal.
The luminosity distance inferred from the \gw observation will be degenerate with the magnification such that the inferred luminosity distance 
\begin{equation}\label{eq:dl_inferred}
 D_{\rm L}^{\rm inferred} = \frac{D_{\rm L}}{\sqrt{|\mu|}}\,. 
\end{equation}
Because of this degeneracy, lensing biases the inferred redshift and thus the source masses. 
Consequently, the binary appears to be closer than it truly is, and it appears to be more massive than it truly is.

\citet{Broadhurst:2018saj,Broadhurst:2020moy,Broadhurst:2020cvm} argued that some of the relatively high-mass LIGO--Virgo events could be strongly lensed \gw[s] from the lower-mass stellar black hole population observed in the electromagnetic bands.
However, the expected strong lensing rates and the current constraints on the merger-rate density,
based on the absence of a detectable \sgwb,
disfavor this interpretation~\citep{Dai:2016igl,Ng:2017yiu,Li:2018prc,Oguri:2018muv,Hannuksela:2019kle,Buscicchio:2020cij,Buscicchio:2020bdq}
compared to the standard interpretation of a genuine unlensed high-mass population~\citep{LIGOScientific:2018jsj,Roulet:2020wyq,GWTC2:rates,Kimball:2020qyd}.
Hence, in the absence of more direct evidence, such as identifying multiple images within LIGO--Virgo data (Sec.~\ref{sec:multiimages}),
it is difficult to support the lensing hypothesis purely based on magnification considerations.
Nevertheless, it is informative to analyze the degree to which the lensed interpretation would change our understanding of the observed sources.

\begin{table*}
\caption{
 \label{tab:mag_lensing_results}
 Inferred properties of selected O3a events under the lensing magnification hypothesis.
}
\begin{center}
\begin{tabular}{lllll}
\hline \hline
 Event name & $m_1$ $[\rm M_\odot]$                             & $m_2$ $[\rm M_\odot]$                              & $z$                       & $\mu$                                     \\ \hline \hline

 GW190425   & $1.42_{-0.12}^{+0.16}$       & $1.27_{-0.15}^{+0.12}$       & $0.3_{-0.1}^{+0.1}$  & $68_{-44}^{+163}$     \\
 GW190426\_152155   & $1.89_{-0.55}^{+0.40}$      & $0.90_{-0.40}^{+0.25}$      & $1.3_{-0.2}^{+0.5}$ & $497_{-272}^{+452}$  \\
 \hline\\
\hline \hline
 Event name & $m_1^{50}$ ($m_1^{65}$) $[\rm M_\odot]$                             & $m_2^{50}$ ($m_2^{65}$) $[\rm M_\odot]$                              & $z^{50}$ ($z^{65}$)                       & $\mu^{50}$ ($\mu^{65}$)                              \\ \hline \hline 
 GW190521   & $43_{-16}^{+6}$ ($55_{-22}^{+9}$)   & $36_{-15}^{+10}$ ($45_{-19}^{+13}$) & $2.5_{-0.7}^{+2.1}$ ($1.8_{-0.5}^{+1.7}$) & $13_{-8}^{+55}$ ($6_{-4}^{+28}$)         \\
 GW190602\_175927  & $42_{-17}^{+7}$ ($48_{-19}^{+14}$)  & $31_{-16}^{+13}$ ($33_{-16}^{+15}$) & $1.4_{-0.5}^{+1.5}$ ($1.1_{-0.4}^{+1.4}$) & $10_{-7}^{+65}$ ($6_{-4}^{+46}$)          \\
 GW190706\_222641  & $39_{-15}^{+10}$ ($42_{-17}^{+17}$) & $29_{-13}^{+12}$ ($29_{-13}^{+13}$) & $1.7_{-0.5}^{+1.8}$ ($1.6_{-0.6}^{+1.7}$) & $5_{-3}^{+26}$ ($4_{-2}^{+22}$)           \\
\hline
\end{tabular}
\end{center}
\tablecomments{Under the hypothesis that the listed events are lensed signals from intrinsically lower-mass binary populations with $\mu>2$, 
this table lists the favored source masses, redshifts, and magnifications 
for the \bns and \nsbh (top) and \bbh (bottom) high-mass events.
For the \bbh[s], two sets of numbers are given for different assumptions about the edge of the \pisn mass gap (a cut at $50 \, \rm M_\odot$ and $65 \, \rm M_\odot$). 
For the \bns[s], we presume that they originate from the Galactic \bns population. 
To interpret the heavy BBHs as lensed signals originating from the assumed lower-mass population, they should be magnified at a moderate magnification $\mu\sim\mathcal{O}(10)$ at $z\sim$~1 to 2. 
The BNS and NSBH events would require extreme magnifications.
}
\end{table*}

Under the strong lensing hypothesis $\lensedhyp $, the \gw would originate from a well-known,
intrinsically lower-mass population,
and the LIGO--Virgo observations have been biased by lensing.
Using such a mass prior, we infer the required magnification and corrected redshift and component masses under $\lensedhyp $.
The posterior distribution of the parameters is \citep{Pang:2020qow}
\begin{equation}
    p(\mu, \vartheta|d,\lensedhyp ) \propto p(d|\vartheta)\, p(\vartheta|\mu,\lensedhyp )\, p(\mu|\lensedhyp ),
\end{equation}
where we distinguish the \emph{apparent} parameters of the waveform received at the detector $\vartheta$, which differ from the intrinsic parameters $\theta$ due to bias by lensing magnification.
Therefore, we can compute the magnification posterior and other parameters by simply re-weighting existing posteriors.

Studies along these lines were already done for the GW190425 \bns event by \citet{Pang:2020qow}
and for the GW190521 \bbh event in \citet{Abbott:2020mjq}.
Here we extend the approach to cover additional interesting O3a events, focusing on two cases:
(i) the (apparently) most massive observed \bbh[s], and (ii) sources with an (apparent) heavy neutron star component. 
In the \bbh case, we take the prior over component masses, $m_1$ and $m_2$, and redshift, $z$ of the source $p(m_1,m_2,z )$
from the power-law \bbh population model used in~\citet{LIGOScientific:2018jsj} for O1 and O2 observations,
with a mass power-law index $\alpha=1$, mass ratio power-law index $\beta_q=0$, and minimum component mass $m_{\rm min}=5\,{\rm M_\odot}$, and assume an absence of \bbh[s] above the \pisn mass gap.
As in the previous GW190521 study~\citep{Abbott:2020mjq}, we consider two different values to account for uncertainties on the edge of the \pisn gap, $m_{\rm max} = (50,65)\,{\rm M_\odot}$.
Such a simple model is adequate for this analysis because our analysis results are most sensitive to the mass cut (highest masses allowed by the prior) and less sensitive to the specific shape of the mass distribution.
For events with an apparent heavy neutron star component,
we assume a Galactic \bns prior following a total mass with a $2.69\,{\rm M}_\odot$ mean and $0.12\,{\rm M}_\odot$ standard deviation~\citep{Farrow:2019xnc}.
In both cases, the magnification could explain the apparent high mass of the events from the LIGO--Virgo observations. 

We assume that the redshift prior $p(z)\propto \tau(z) \mathrm{d}V_\mathrm{c}/\mathrm{d}z$, where the optical depth of lensing by galaxies or galaxy clusters $\tau(z)\propto D_{\rm c}(z)^3$ \citep{Haris:2018vmn}. 
The redshift dependence of the optical depth is approximately the same for both galaxies and galaxy clusters, while the overall scaling can change~\citep{Fukugita:1991yz}. 
We use the lensing prior \mbox{$p(\mu|\lensedhyp )\propto \mu^{-3}$} \citep{Blandford:1986zz}
with a lower limit $\mu>2$ appropriate to strong lensing~\citep{Ng:2017yiu}.
This prior is appropriate when we are in the high-magnification, strong lensing limit, i.e.\ assuming that the observed masses are highly biased. 
We do not consider weak lensing, which does not produce multiple images
and would require expanded future \gw data sets to study~\citep{Mukherjee:2019wcg, Mukherjee:2019wfw}.

We analyze all O3a \bbh events with the primary mass above $50 \, \rm M_\odot$ at $90\%$ probability using the Bayesian inference posterior samples released with GWTC-2~\citep{gwosc:gwtc2,GWTC2}.
Moreover, we analyze GW190425, a high-mass \bns~\citep{GW190425},
and GW190426\_152155, a low-significance potential \nsbh event~\citep{GWTC2} which was investigated as a possible lensed BNS event~\citep{2019GCN.24250....1S}. 
We use the results for the \IMRP waveform~\citep{Hannam:2013oca,Bohe:PPv2} for most of the events.
For GW190521, where higher-order multipole moments are important to include in the analysis~\citep{Abbott:2020mjq},
we adopt the \NRSur waveform~\citep{Varma:2019csw} results as in~\citet{Abbott:2020tfl}.
Furthermore, for GW190425~\citep{GW190425}, we use the \IMRPNRT~\citep{Dietrich:2018uni} low-spin samples.
Results are summarized in Table~\ref{tab:mag_lensing_results}.

To interpret the heavy \bbh[s] as lensed signals originating from the assumed lower-mass population, they should be magnified at a moderate magnification $\mu \sim 10$ at $z \sim 1\textendash2$.
Depending on the lens model, this magnification may imply a moderate chance of an observable multi-image counterpart as events closer to the caustic curves experience more substantial magnifications. 
Consequently, they often produce events with similar magnification ratios and shorter time delays (comparable magnifications and shorter time delays can be derived from the lens's symmetry, although if lensing by substructures or microlenses is present, the magnifications between images can differ even in the high-magnification limit).
However, we could not identify any multi-image counterparts for any of the high-mass events in our multiple image search (Sec.~\ref{sec:multiimages}). 

The \bns and \nsbh events, on the other hand, would require extreme magnifications ($68_{-44}^{+163}$ and $497^{+452}_{-272}$, respectively) to be consistent with the Galactic \bns distribution.
At these magnifications, we would expect the source to be close to a caustic, and therefore it may be possible that the presence of microlenses would produce observable effects~\citep{Diego:2019lcd,Diego:2019rzc,Pagano:2020rwj,Mishra:2021abc}.
Moreover, the event would likely be multiply imaged~\citep{Blandford:1986zz,Oguri:2018muv}.
A more detailed follow-up study to quantify the likelihood of multiple images and microlensing could produce more stringent evidence for the lensing hypothesis for these events. 
We will briefly comment on these events in the context of multi-image and microlensing results in the sections that follow.

At this stage, we cannot set robust constraints on the lensing hypothesis based on the magnification alone. 
Moreover, as detailed in the following section, we have also not found any other clear evidence to indicate that these \gw events are lensed.
The prior lensing rate disfavors the lensing hypothesis for most standard binary population and lens models, as discussed in Sec.~\ref{sec:statistics}. However, if other \bbh formation channels exist that produce an extensive number of mergers at high redshift, the lensing rates can change. 
In the future, more quantitative constraints could be set by connecting the inferred magnifications with lens modeling to make predictions for the appearance of multiple images or microlensing effects.

\section{Search for multiple images} 
\label{sec:multiimages}
In addition to magnification, strong lensing can produce multiple images of a single astrophysical event.
These multiple images appear at the \gw detectors as repeated events. 
The images will differ in their arrival time and amplitude~\citep{Wang:1996as,Haris:2018vmn,Hannuksela:2019kle,Li:2019osa,McIsaac:2019use}.
The sky location is the same within the localization accuracy of \gw detectors, given that the typical angular separations are of the order of arcseconds.
Additionally, lensing can invert or Hilbert transform the image~\citep{Dai:2017huk,Ezquiaga:2020gdt}, introducing a frequency-independent phase shift.
This transformation depends on the image type, set by the lensing time delay at the image position: Type-I, II, and III correspond to a time-delay minimum, saddle point, and maximum, respectively~\citep{Ezquiaga:2020gdt}.

The multiply imaged waveforms $\{\tilde{h}_j^L\}$ of a single signal $\tilde{h}$ then satisfy~\citep{Dai:2017huk, Ezquiaga:2020gdt}
\begin{equation} \label{eq:multiple_image_waveform}
 \tilde{h}_j^L(f; \theta, \mu_j, \Delta t_j, \Delta \phi_j) = \sqrt{|\mu_j|} \tilde{h}(f; \theta, \Delta t_j) \exp\big(i \, {\rm sign}(f) \Delta \phi_j \big)\,,
\end{equation}
where $\sqrt{|\mu_j|}$ is the lensing magnification experienced by the image $j$
and $\Delta \phi_j=-\pi n_j /2$ is the Morse phase, with index $n_j=0,1,2$ for Type-I, II, and III images.
$\tilde{h}(f; \theta, \Delta t_j)$ is the original (unlensed) waveform before lensing, but evaluated as arriving with a time delay $\Delta t_j$. 
The multi-image hypothesis then states that most parameters measured from the different lensed images of the same event are consistent. 

The relative importance of different parameters for the overall consistency under the multi-image hypothesis will vary for different events. 
For example, the sky localization match will have greater relevance for well-localized, high-\snr events.
Similarly, the overlap in measured chirp mass $(1+z)\Mc = (1+z)\,(m_{1}m_{2})^{3/5}/(m_{1}+m_{2})^{1/5}$ will be more significant when the uncertainty in that parameter is lower, although in this case the underlying astrophysical mass distribution will play a key role.
The similarities in other parameters such as mass ratios or spins will be more important when they depart from the more common astrophysical expectations.
Evidence of strong lensing could also be acquired with a single Type-II (saddle point) image if the induced waveform distortions in the presence of higher modes, precession, or eccentricity are observed \citep{Ezquiaga:2020gdt}.
Such evidence is unlikely to be observed without next-generation detectors~\citep{Wang:2021kzt}.

In this section, we perform three distinct but related analyses. 
First, we test the lensed multi-image hypothesis by analyzing, for all pairs of O3a events from GWTC-2,
the overlap of posterior distributions previously inferred for the individual events.
This allows us to set ranking statistics to identify an initial set of candidates for lensed multiple images.
We perform a more detailed joint-\pe analysis for these most promising pairs, considering all potential correlations in the full parameter space and the image type.
This joint analysis provides a more solid determination of the lensing probability for a given \gw pair.
Finally, we search for additional sub-threshold candidates that could be multiply imaged counterparts to the previously considered events:
some counterpart images can have lower relative magnification compared with the primary image and/or fall in times of worse detector sensitivity or antenna patterns,
and hence may not have passed the detection threshold of the original broad searches.
According to the predictions of the expected lensing time delays and the rate of galaxy and galaxy cluster lensing~\citep{Smith:2017mqu,Oguri:2018muv,Dai:2020tpj},
we expect it to be less likely for counterpart images to the O3a events to be detected in observing runs O1 or O2. 
Relative lensing rates for galaxies and clusters are given in Table~\ref{tab:lensedrates0_min}.
Thus, we only search for multiple images within O3a itself.

Previous studies have also searched for multiple images in the O1--O2 catalog GWTC-1~\citep{Hannuksela:2019kle,Broadhurst:2019ijv,Li:2019osa,McIsaac:2019use,Dai:2020tpj,Liu:2020par}.
The first search for \gw lensing signatures in O1 and O2 focused on the posterior overlap of the masses, spins, binary orientation and sky positions~\citep{Hannuksela:2019kle} and the consistency of time delays with expectations for galaxy lenses, but found no conclusive evidence of lensing.
The search did uncover a candidate pair GW170104--GW170814 with a relatively high Bayes factor of $\gtrsim 200$.
Still, this study disfavored the candidate due to its long time delay and the low prior probability of lensing.
In parallel, \citet{Broadhurst:2019ijv} suggested that the candidate pair GW170809--GW170814 could be lensed, but this claim is disfavored by more comprehensive analyses~\citep{Hannuksela:2019kle,Liu:2020par}.
Both \citet{Li:2019osa} and \citet{McIsaac:2019use} performed searches for sub-threshold counterparts to the GWTC-1 events, identifying some marginal candidates but finding no conclusive evidence of lensing.
More recently, \citet{Dai:2020tpj} and \citet{Liu:2020par} searched for lensed \gw signals including the analysis of the lensing image type, which can be described through the Morse phases, $\Delta \phi_j$ in Eq.~\eqref{eq:multiple_image_waveform}.
These analyses have revisited the pair GW170104--GW170814 and demonstrated that the Morse phase is consistent with the lensed expectation but would require Type-III (time-delay maximum) images, which are rare from an observational standpoint.
\citet{Dai:2020tpj} also pointed out that a sub-threshold trigger, designated by them as GWC170620, is also consistent with coming from the same source.
However, the required number and type of images for this lens system make the interpretation unlikely given current astrophysical expectations.
Also, two same-day O3a event pairs
(on 2019 May 21 and 2019 August 28)
have already been considered elsewhere,
but were both ruled out due to vanishing localization overlap~\citep{Singer:2019vjs,Abbott:2020mjq}.

    \subsection{Posterior-overlap analysis}
    \label{sec:posterioroverlap}
    As a consequence of degeneracies in the measurements of parameters, the lensing magnification can be absorbed into the luminosity distance (Sec.~\ref{sec:individualevents}), the time delay can be absorbed into the time of coalescence,
and, when the radiation is dominated by $\ell=|m|=2$ multipole moments,
the phase shifts introduced by lensing (the Morse phases) can be absorbed into the phase of coalescence.
The multi-image hypothesis then states that all other parameters except the arrival time, luminosity distance, and coalescence phase are the same between lensed events, and thus there should be extensive overlap in their posterior distributions, even if those have been inferred without taking lensing into account.

Therefore, we use the consistency of \gw signals detected by LIGO and Virgo to identify potential lensed pairs.
Following~\cite{Haris:2018vmn}, we define a ranking statistic $\posstatistic$ to distinguish candidate lensed pairs from unrelated signals:
\begin{equation}
\posstatistic = \int \mathrm{d} \Theta~ \frac{p(\Theta|d_1)~p(\Theta|d_2)}{p(\Theta)}~,
\label{eq:Bayes_fac}
\end{equation}
where the parameters $\Theta$ include the redshifted masses $ (1+z) m_{1,2}$,
the dimensionless spin magnitudes $\chi_{1,2}$,
the cosine of spin tilt angles $\theta_{1,2}$,
the sky location $(\alpha, \sin \delta)$,
and the cosine of orbital inclination $\theta_{JN}$, 
but they do not include the full 15-dimensional set of parameters $\Theta$ to ensure the accuracy of the~\kde[s] that we use to approximate the posterior distributions $p(\Theta|d_{1,2})$ for each event when evaluating Eq.~\eqref{eq:Bayes_fac}. Here, 
$p(\Theta)$ denotes the prior on $\Theta$. 

The accuracy of the~\kde approximation was demonstrated in \citet{Haris:2018vmn} through receiver operating characteristic curves with simulated lensed and unlensed \bbh events. 
To improve the accuracy further, we compute the sky localization ($\alpha$,$\delta$) overlap separately from other parameters and combine it with the overlap from the remaining parameters. 
Splitting the two overlap computations is justified because the posterior correlations of ($\alpha$,$\delta$) with other parameters are minimal.

We use posterior samples~\citep{gwosc:gwtc2} obtained using the \linf software package~\citep{Veitch:2014wba} with the \IMRP waveform model~\citep{Hannam:2013oca,Bohe:PPv2} for most of the events.
However, for GW190521, we use \NRSur~\citep{Varma:2019csw} posteriors, and for GW190412 and GW190814 we use \IMRPHM~\citep{Khan:2019kot} posteriors. The prior
$p(\Theta)$ is chosen to be uniform in all parameters.
The component mass priors have the bound $(2\textendash200\,\Msun)$.
Equation~\eqref{eq:Bayes_fac} then quantifies how consistent a given event pair is with being lensed.
In our analysis, we omit the \bns event GW190425~\citep{GW190425} because it was detected at relatively low redshift, and hence we expect the probability of it being lensed to be very small.

In addition to the consistency of the frequency profile of the signals (as measured by the posterior overlap),
the expected time delays $\Delta t$ between lensed images follow a different distribution than for pairs of unrelated events. Following~\citet{Haris:2018vmn}, we define
\begin{equation}
\mathcal{R}^{\mathrm{gal}}= \frac{p(\Delta t |\lensedhyp )}{p(\Delta t | \notlensedhyp )},
\label{eq:td_odd_ratio}
\end{equation}
where $p(\Delta t |\lensedhyp )$ and $p(\Delta t |\notlensedhyp )$ are the prior probabilities of the time delay $\Delta t$ under the strongly lensed and unlensed hypotheses, respectively.
Here $p(\Delta t |\notlensedhyp )$ is obtained by assuming that the \gw events follow a Poisson process.
We use a numerical fit to the time-delay distribution $p(\Delta t|\lensedhyp )$ obtained in Sec.~\ref{sec:statistics} for the SIS galaxy lens model, with a merger rate density given by $R_{\min}$ in Eq.~\eqref{app:eq:rate_rp}.
Equation \eqref{eq:td_odd_ratio} provides another ranking statistic to test the lensing hypothesis, based on the time delay, though subject to some astrophysical uncertainties (see discussion in Sec.~\ref{sec:statistics}).
The time-delay distribution does not include galaxy cluster lenses, which may be responsible for long time delays of several months or more. 
We also do not model detector downtime, but we expect the different contributions to the time delay to average out across a longer time period.

\begin{figure}
	\centering
	\includegraphics[width=0.5\textwidth]{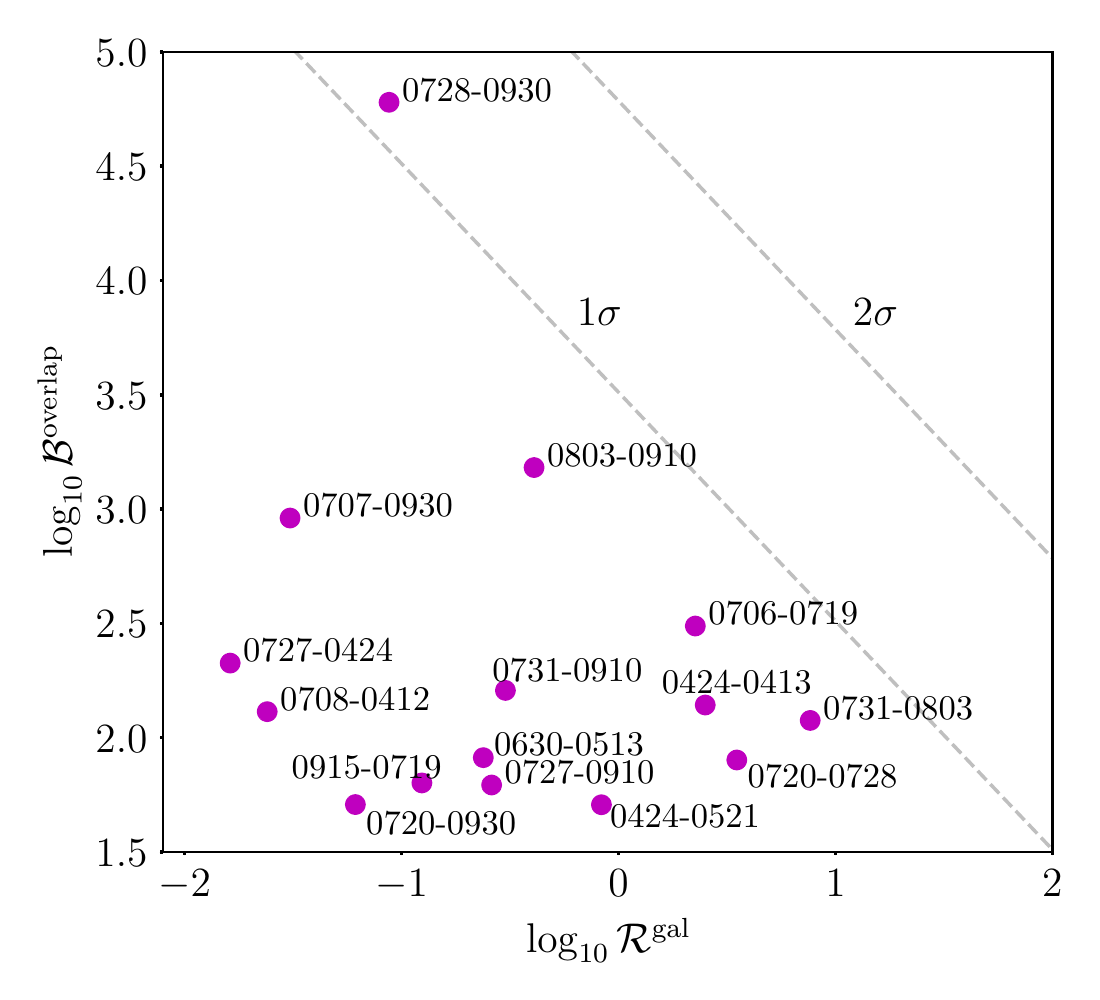}
	\caption{Scatter plot of the ranking statistics $\log_{10} \posstatistic$ and $\log_{10} \mathcal{R}^{\mathrm{gal}}$ for a subset of event pairs that have both $\posstatistic > 50 $ and $\mathcal{R}^{\mathrm{gal}} > 0.01$.
	The dashed lines denote the significance levels of the combined ranking statistics (in terms of Gaussian standard deviations), obtained by simulating unlensed event pairs in Gaussian noise matching the O3a sensitivity of the LIGO--Virgo network.
	We identify several high $\posstatistic > 50$ candidates, which we follow up on with a detailed joint-\pe analysis.
    We have used abbreviated event names, quoting the last 4 digits of the date identifier (see Table~\ref{table:pe-blu} for full names).
	\label{fig:posterior_overlap_blu_scatter_plot}
	}
\end{figure}

To estimate the significance of the combined ranking statistic, $\log_{10}(\posstatistic \times \mathcal{R}^{\mathrm{gal}} )$ computed for O3a event pairs, we perform an injection campaign.
For the injection campaign, we sample component masses $m_{1,2}$ from a power-law distribution \citep{Abbott:2016nhf} in the range $(10\textendash50\, \Msun)$.
We assume that the redshift distribution follows population synthesis simulations of isolated binary evolution  \citep{Belczynski:2005mr,Belczynski:2010tb,Dominik:2013tma,Marchant:2018kun,Eldridge:2018nop,Neijssel:2019irh,Boco:2019teq,Santoliquido:2020axb}; in particular, for illustration purposes, we show results using the redshift evolution from  \citet{Belczynski:2016obo,Belczynski:2016jno}, but for the local universe that we look at ($z<2$), other models produce qualitatively similar results.  All other parameters are sampled from uninformative prior distributions \citep{Haris:2018vmn}.
We inject the simulated signals into Gaussian noise with O3a representative spectra for a LIGO\textendash Virgo detector network. 
We compute $\posstatistic$ and $\mathcal{R}^{\mathrm{gal}}$ for all possible pairs in the injection set to obtain the false-alarm probability for one pair
$\text{FAP}^{\mathrm{pair}}(x)$ 
at different levels $x$  of combined statistics
by counting the number of simulated pairs with
$\log_{10}(\posstatistic \times \mathcal{R}^{\mathrm{gal}} ) > x$.
Then the probability of at least one of the $N$ event pairs in GWTC-2 to cross the threshold can be estimated as
$\text{FAP}^{\mathrm{cat}}(x) = 1-[\text{FAP}^{\mathrm{pair}}(x)]^N$.
We then obtain the $\sigma$ levels of significance shown in Fig.~\ref{fig:posterior_overlap_blu_scatter_plot}
by assuming $\text{FAP}^{\mathrm{cat}}(x)$ follows the complementary error function.

In Fig.~\ref{fig:posterior_overlap_blu_scatter_plot} we show the scatter plot of $\log_{10} \posstatistic$ and $\log_{10} \mathcal{R}^{\mathrm{gal}}$ for the O3a event pairs that have high combined ranking statistic.
The dashed lines represent different significance levels as obtained from the simulations.
The event pair GW190728\_064510--GW190930\_133541 gives the highest combined ranking statistic,
$\log_{10} (\posstatistic \times \mathcal{R}^{\mathrm{gal}}) = 3.6$; however, as can be seen from Fig.~\ref{fig:posterior_overlap_blu_scatter_plot}, its significance is above 1$\sigma$ (68$\%$) but much below the 2$\sigma$ (95$\%$) significance level.

To follow up on the most promising event pairs with the more detailed joint-\pe analysis in the next section, we make a selection based on just the posterior overlap ranking statistic, $\posstatistic$, rather than the combined ranking statistic, $\posstatistic \times \mathcal{R}^{\mathrm{gal}}$,
because $\mathcal{R}^{\mathrm{gal}}$ depends strongly on the lens model.
That is, we do not rule out any candidates based on $\mathcal{R}^{\mathrm{gal}}$.
Our aim in the next section is to understand the high $\posstatistic$ event pairs in greater detail without resorting to any specific lens model.
We thus select the most promising event pairs from Fig.~\ref{fig:posterior_overlap_blu_scatter_plot}, i.e., those with $\posstatistic >50$, and carry out the joint-\pe analysis in the next section.
The 19 selected pairs are listed in Table~\ref{table:pe-blu}.

    \subsection{Joint parameter estimation analysis}
    \label{sec:jointpe}
    Here we follow up on the most significant pairs of events from the posterior-overlap analysis with a more detailed but more computationally demanding joint-\pe analysis.
The benefit of this analysis is that it allows for more stringent constraints on the lensing hypothesis by investigating potential correlations in the full 
parameter space of \bbh signals, instead of marginalizing over some parameters.
Moreover, it also includes a test for the lensing image type by incorporating lensing phase information.

We perform our analysis using two independent pipelines, a \linf{}-based pipeline~\citep{Liu:2020par} and a \bilby-based pipeline \citep[\hanabi;][]{Lo:2021nae}, giving us additional confidence in our results.
Unlike the posterior-overlap analysis, the joint-\pe analysis does not start from existing posterior samples.
Instead, we start the inference directly using the detector strain data. 
In both pipelines, we follow the same data selection choices
(calibration version, available detectors for each event, and noise subtraction procedures)
as in the original GWTC-2 analysis~\citep{GWTC2},
with special noise mitigation steps (glitch subtraction and frequency range limitations) taken for some events as listed in Table V of that paper.
However, the two pipelines use different waveform models.
In this section, we first describe how we quantify the evidence for the strong lensing hypothesis, then detail the two pipelines and finally present the results. 

\subsubsection{The coherence ratio and the Bayes factor}

There will be three types of outputs for the joint-\pe analysis.
First, we compute a coherence ratio $\cohratio$, which is the ratio of the lensed and unlensed evidences, neglecting selection effects and using default priors in the joint-\pe inference.
We treat this as a ranking statistic, which quantifies how consistent two signals are with the lensed hypothesis. 
Large coherence ratios indicate that the parameters of the \gw[s] agree with the expectations of multiple lensed events. This occurs, for example, when the masses and sky localization coincide.  
However, the coherence ratio does not properly account for the possibility that the parameters overlap by chance. 

The likelihood that \gw{} parameters overlap by chance sensitively depends on the underlying population of sources and lenses. 
For example, if there existed formation channels that produced \gw[s] with similar frequency evolutions (as expected of lensing), the likelihood of an unlensed event mimicking lensing would increase substantially. 
Thus, we introduce a second output, the population-weighted coherence ratio $\cohratio|_\mathrm{pop}$, which incorporates prior information about the populations of \bbh[s] and lenses. The value of $\cohratio|_\mathrm{pop}$ is subject to the choice of both the \bbh and lens models. 

Similarly, the probability that two signals agree with the multiple-image hypothesis is altered through selection effects, as some masses and sky orientations are preferentially detected. 
Thus, we also include the selection effects, which gives us our final output, the Bayes factor $\Blu$. 
The $\Blu$ quantifies the evidence of the strong lensing hypothesis for a given detector network and population model. 
For the full derivations and detailed discussion on the difference between the coherence ratio and the Bayes factor, see~\cite{Lo:2021nae}.

\subsubsection{\linf{}-based pipeline}
\label{sec:jointpe-linf}

For the \linf{}-based pipeline, we adopt the method presented by \citet{Liu:2020par}, which was first used for analyzing pairs of events from GWTC-1~\citep{GWTC1}.
\linfnest~\citep{Veitch:2014wba} implements nested sampling~\citep{Skilling:2006gxv}, which can compute evidences without explicitly carrying out the high-dimensional integral while sampling the posteriors.
The \linf{}-based pipeline uses the \IMRD waveform~\citep{Husa:2015iqa, Khan:2015jqa}, which is a phenomenological model that includes the inspiral, merger, and ringdown phases but assumes non-precessing binaries and only $\ell = |m|=2$ multipole radiation.
This is motivated by the fact that most events detected so far are well described by the dominant multipole moment~\citep{GWTC1,GWTC2}. 
Higher-order multipole moments, precession, or eccentricity could lead to non-trivial changes to the waveform for Type-II images,
but such waveforms cannot currently be used with this pipeline. 
For a discussion of the events within GWTC-2 displaying measurable higher-order multpole moments or precession, see Appendix A of~\citet{GWTC2}.

As in the posterior-overlap analysis, we expect observed, lensed \gw[s] to share the same parameters for the redshifted masses, spins, sky position, polarization angle and inclination, $\{(1+z)m_1, (1+z)m_2, \chi_1, \chi_2, \alpha, \delta, \psi, \theta_{JN} \}$. 
Hence, we force these parameters to be identical under the lensing hypothesis.
For the unlensed hypothesis, we sample independent sets of parameters for each event.
This is equivalent to performing two separate nested sampling runs and then combining their evidence. 
In total, \linf{} samples in an 11-dimensional parameter space and provides $\cohratio$ as the output.

We sample the apparent luminosity distance of the first event $D_L^1$ and the relative magnification $\mu_\mathrm{r}$ \citep{Wang:1996as} instead of the luminosity distance of the second event $D_L^2$, using the relation $\sqrt{\mu_\mathrm{r}} = D_L^1/D_L^2$.
Since our waveform only includes the dominant $\ell = |m|=2$ multipole moments,
the lensing Morse phase is modeled by discrete shifts in the coalescence phase $\phi_\mathrm{c}$ by an integer multiple of $\pi/4$ \citep[with relation to the lensing phase shift $\Delta\phi=2\Delta\phi_\mathrm{c}$,][]{Dai:2017huk, Ezquiaga:2020gdt}.
Thus, we consider all possible relative shifts $\Delta\phi_\mathrm{c} \in \{0, \pi/4, \pi/2, 3\pi/4\}$ between two \gw signals.

We set a uniform prior in $\log[(1+z)m_{1}]$ and $\log[(1+z)m_{2}]$ for both the lensed and unlensed hypothesis. The minimum and maximum component masses are respectively $3\,\Msun$ and $330\,\Msun$, with a minimum mass ratio of $q=m_2/m_1=0.05$.
This choice reduces the prior volume by $10^2-10^3$ compared to the uniform prior used in GWTC-2 \citep[see][ for discussion]{Liu:2020par}.
For the other parameters, the prior for the luminosity distance is $p(D_\mathrm{L})\propto D_\mathrm{L}^2$ up to $20$ Gpc, while the spins are taken to be parallel to the dimensionless orbital angular momentum with a uniform prior on the $z$ components between $-0.99$ (anti-aligned) and $+0.99$ (aligned).

\subsubsection{The \hanabi{} pipeline}
\label{sec:jointpe-hanabi}
The \hanabi{} pipeline, on the other hand, adopts a hierarchical Bayesian framework that models the data generation process under the lensed and the unlensed hypothesis.
This pipeline uses the \IMRXPHM waveform \citep{Pratten:2020ceb}, which models the full inspiral--merger--ringdown for generic precessing binaries including both the dominant and some sub-dominant multipole moments.
Therefore, the parameter space of \hanabi{} enlarges to 15 dimensions. 

\hanabi{} differs from the \linf{}-based pipeline in the treatment of the Morse phase. Here the lensing phase is directly incorporated in the frequency-domain waveform, accounting for any possible distortion of Type-II images \citep{Dai:2016igl, Ezquiaga:2020gdt, Lo:2021nae}.
Moreover, the lensed probability is computed by considering all possible combinations of image types with a discrete uniform prior~\citep{Lo:2021nae}.
For this reason, \hanabi{} only produces one evidence per pair, and not one for each discrete phase difference as the \linf{}-based pipeline.
Unlike the \linf{}-based pipeline, \hanabi{} samples the observed masses in a uniform distribution.
The mass ranges are different for each event pair, but an overall reweighting is applied later (see below).
The rest of the prior choices for the intrinsic parameters are the same as for the \linf{}-based pipeline with the addition of a discrete uniform prior on the Morse phase and isotropic spin priors.

In addition to computing the joint-PE coherence ratio, \hanabi{} also incorporates prior information about the lens and \bbh populations, as well as selection effects.
In particular, the \bbh population is chosen to follow a \textsc{Power Law + Peak model} in the primary mass following the best-fit parameters in \citet{GWTC2:rates}.
Similarly, the secondary mass is fixed to a uniform distribution between the minimum and the primary mass. \hanabi{} also uses an isotropic spin distribution and merger rate history following Model A in Sec. \ref{sec:statistics}. 
The lens population is modeled by the optical depth described in \citet{Hannuksela:2019kle} and a magnification distribution $p(\mu)\propto \mu^{-3}$ for $\mu\geq2$. 
\hanabi{} is thus able to output $\cohratio$, $\cohratio|_\mathrm{pop}$ and $\Blu$.
However, \hanabi{} does not include any preference for a particular type of image, i.e., \hanabi{} uses a discrete, uniform prior for the Morse phase shift $\Delta \phi_j$.

\begin{table*}
\caption{
 \label{table:pe-blu}
 Summary of joint-PE results for event pairs in O3a.
}
\begin{tabular*}{\textwidth}{c @{\extracolsep{\fill}} lccccr }
\hline
	& & & $\log_{10}(\cohratio)$ & $\log_{10}(\left.\cohratio\right|_\mathrm{pop})$ & $\log_{10}(\Blu)$\\ 
	Event 1 & Event 2 &  $\log_{10}~\mathcal{R}^\mathrm{gal}$ &\linf{} & \hanabi{} & \hanabi{} & \\
	& & & ($\Delta\phi$: $0$, $\pi/2$, $\pi$, $3\pi/2$) & & & \\[1ex]
	\hline
	GW190412 & GW190708\_232457   & $-1.6$ & $\left( +1.0, -9.7, -22.8, -4.4 \right)$  & $-6.6$ & $-9.7$  \\[1ex]
	\textbf{GW190421\_213856} & \textbf{GW190910\_112807}   & $-$ & $\left( \mathbf{+4.5}, +2.5, -1.5, -0.0\right)$ & $-0.7$ & $\mathbf{-3.8}$  \\[1ex]
	\textbf{GW190424\_180648} & \textbf{GW190727\_060333}   & $-1.8$ & $\left( \mathbf{+4.9}, +0.0, +1.1, \mathbf{+4.0}\right)$ & $-0.8$ & $\mathbf{-3.9}$ \\[1ex]
	\textbf{GW190424\_180648} & \textbf{GW190910\_112807}  & $-$ & $\left( +2.5, \mathbf{+4.7}, \mathbf{+4.3}, +1.6\right)$ & $-0.8$ & $\mathbf{-3.9}$  \\[1ex]
	\textbf{GW190513\_205428} & \textbf{GW190630\_185205}  & $-0.6$  & $\left( +0.8, \mathbf{+4.3}, -1.9, -6.5\right)$ & $-2.4$ & $\mathbf{-5.5}$  \\[1ex]
	GW190706\_222641 & GW190719\_215514  & $+0.4$  & $\left( +2.4, +2.4, -0.0, -0.5\right)$ & $-0.3$ & $-3.4$  \\[1ex]
	GW190707\_093326 & GW190930\_133541  & $-1.5$  & $\left( -4.6, -4.3, -3.5, -4.1\right)$ & $-9.4$ & $-12.5$ \\[1ex]
	\textbf{GW190719\_215514}  & \textbf{GW190915\_235702} & $-0.9$ & $\left( +3.5, -2.1, -0.1, \mathbf{+4.1}\right)$  & $-0.7$ & $\mathbf{-3.8}$ \\[1ex]
	GW190720\_000836 & GW190728\_064510 & $+0.5 $ & $\left( -1.4, -0.9, -4.5, -5.4\right)$  & $-6.7$ & $-9.8$ \\[1ex]
	GW190720\_000836 & GW190930\_133541 & $-1.2$  & $\left( -3.5, -2.8, -3.9, -3.9\right)$  & $-9.2$ & $-12.3$  \\[1ex]
	GW190728\_064510 & GW190930\_133541 & $-1.1$  & $\left( -3.6, -2.5, -3.1, -2.9\right)$  & $-8.5$ & $-11.6$  \\[1ex]
	GW190413\_052954 & GW190424\_180648 & $+0.4 $ &  $\left( +0.6, -0.9, +0.4, -0.0\right)$  & $-1.6$ & $-4.7$  \\[1ex]
	\textbf{GW190421\_213856} & \textbf{GW190731\_140936} & $-2.1$   &  $\left( +3.1, -1.9, +2.5, \mathbf{+5.2}\right)$  & $-0.2$ & $\mathbf{-3.3}$  \\[1ex]
	\textbf{GW190424\_180648} & \textbf{GW190521\_074359} & $-0.1$   &  $\left( +1.3, +3.8, +3.7, \mathbf{+4.4}\right)$ & $-2.0$ & $\mathbf{-5.1}$  \\[1ex]
	\textbf{GW190424\_180648} & \textbf{GW190803\_022701} & $-2.1$   & $\left( \mathbf{+4.2}, +1.9, +2.6, +3.1\right)$  & $-1.0$ & $\mathbf{-4.1}$  \\[1ex]
	GW190727\_060333 & GW190910\_112807 &	$-0.6$   &  $\left( +1.8, +3.3, +3.7, +3.4\right)$  & $-1.4$ & $-4.5$  \\[1ex]
	\textbf{GW190731\_140936} & \textbf{GW190803\_022701} & $+0.9 $  & $\left( \mathbf{+4.1}, +3.2, +2.2, +3.4\right)$   & $-0.9$ & $\mathbf{-4.0}$ \\[1ex]
	\textbf{GW190731\_140936} & \textbf{GW190910\_112807} & $-0.5$   & $\left( +0.1, \mathbf{+4.5}, +0.8, -7.2\right)$  & $-1.2$ & $\mathbf{-4.3}$  \\[1ex]
    \textbf{GW190803\_022701} & \textbf{GW190910\_112807} & $-0.4$   & $\left( \mathbf{+4.0}, \mathbf{+5.5}, \mathbf{+4.7}, +2.6\right)$  & $-0.1$ & $\mathbf{-3.2}$   \\[1ex]
\hline
\end{tabular*}
\tablecomments{
We select those events with posterior overlap ranking statistic larger than $50$.
For each pair of events presented in the first two columns,
the third column lists the time-delay ranking statistic $\mathcal{R}^\mathrm{gal}$ as described in Section \ref{sec:posterioroverlap}.
The next column gives the coherence ratio of the lensed/unlensed hypothesis $\cohratio$
obtained with the \linf{}-based pipeline, including the results for the four possible lensing phase difference $\Delta\phi=2 \Delta\phi_\mathrm{c}$.
We highlight in boldface those pairs with $\log_{10}(\cohratio)>4$ for at least one Morse phase shift.
The fifth and sixth columns correspond to the \hanabi{} results for the population-weighted coherence ratio $\cohratio|_\mathrm{pop}$ and the Bayes factor $\Blu$.
All quantities are given in $\log_{10}$.
All high coherence ratio events display a small Bayes factor when including the population priors and selection effects.
For the pairs GW190421\_213856--GW190910\_112807 and GW190424\_180648--GW190910\_112807,
the time delays between events are larger than what we expect for galaxy lenses in our simulation,
and thus \mbox{$\mathcal{R}^\mathrm{gal} = 0$}. 
}
\end{table*}

\subsubsection{Results}
Within the O3a events, 
the \linf{}-based pipeline finds 11 pairs with $\log_{10}(\cohratio)>4$, indicating high parameter consistency. 
We have checked that the results of the \linf{}-based pipeline are qualitatively consistent with those from \hanabi.
This reinforces our previous argument that the shift in the coalescence phase is a good approximate description of the lensing Morse phase given that in the present catalog most events are dominated by the $\ell=|m|=2$ multipole moments.
However, because of the pair-dependent prior choices of \hanabi{}, we do not present its raw $\cohratio$ results in Table~\ref{table:pe-blu}.

We then include our prior expectation on the properties of the lensed images (derived from our \bbh{} and lens population priors) and selection effects when computing the population-weighted \hanabi coherence ratio and the Bayes factors $\Blu$.
The results are summarized in Table~\ref{table:pe-blu}.
The event pair GW190728\_064510--GW190930\_133541,
which seemed the most promising from the overlap analysis in Sec.~\ref{sec:posterioroverlap},
is disfavored by both joint-\pe pipelines.
After the inclusion of the population prior and selection effects,
none of the event pairs display a preference for the lens hypothesis ($\log_{10}\Blu<0$). 

The population-weighted coherence ratio and the Bayes factor are subject to the \bbh{} and lens model specifications.
The population properties are not inferred taking into account the possibility of lensing.
This introduces an inevitable bias, but it can be justified \emph{a posteriori} to be a good approximation given the expected low rate of strong lensing.
Additionally, the population properties include significant uncertainties in the hyper-parameter estimates and presume a population model.
In any case, to quantify this intrinsic uncertainty in the modeling, we consider different choices for the mass distribution and merger rate history. 
Varying the maximum \bbh{} mass and the redshift evolution of the merger rate using the $R_{\rm min}(z)$ and $R_{\rm max}(z)$ of Model A in Sec. \ref{sec:statistics}, we find that the strong lensing hypothesis is always disfavored. 
While these results are subject to assumptions on prior choices, our results are sufficient to reject the strong lensing hypothesis: Even if other prior choices favored the lensing hypothesis, the evidence would at best be inconclusive. 

The impact of selection effects is considerable.
Among other reasons, this is because present \gw{} detectors preferentially observe higher mass events \citep{Fishbach:2017zga}, making coincidences in observed masses more probable.
Along the same lines, given the specific antenna patterns of the current network of detectors,
\gw{} events are preferentially seen in specific sky regions with characteristic elongated localization areas \citep{Chen:2016luc},
which favors the overlap between different events.

We also reanalyze the GW170104--GW170814 event pair
in the O2 data previously studied by~\citet{Dai:2020tpj,Liu:2020par}.
Using the \linf{}-based pipeline, \citet{Liu:2020par} found that the coherence ratio, including selection effects associated with the Malmquist bias \citep{1922MeLuF.100....1M}, is $\log_{10}(\cohratio)\approx 4.3$ for a $\pi/2$ coalescence phase shift. 
However, when including together population and selection effects with \hanabi{},
we find that the evidence drastically reduces to a Bayes factor of $\log_{10}(\Blu) \approx -2.0$.

In addition to the Bayes factor, it is important to contrast the recovered number of candidate lensed pairs and their properties with astrophysical expectations.
In Sec. \ref{sub:lensing_rate_strong} we found that the relative rate of \gw{} events with at least two strongly lensed images above the detection threshold is below $\sim 1.3\times 10^{-3}$ for all considered \bbh population models. 
Thus, the lensing rate estimates significantly disfavor the lensing hypothesis \emph{a priori}; even a moderate Bayes factor would not by itself yet make a compelling case for strong lensing. 
Additionally, the type of images, arrival times, and magnifications provide additional information on the lensing interpretation's plausibility.
For example, a quantification of the time-delay prior can be computed by multiplying the coherence ratio by $\mathcal{R}^{\mathrm{gal}}$. 
However, our final conclusions do not depend on the prior information about the lensing time delays or the prior odds against lensing:
the prior lensing knowledge further disfavors the strong lensing hypothesis, but we did not use it to rule out any candidates.

Although we do not find evidence of strong lensing, future electromagnetic follow-up of the candidates could allow for independent support for the hypothesis 
if we identified a lensed counterpart galaxy to these events~\citep{Sereno:2011ty,Smith:2017mqu,Smith:2018gle,Smith:2019dis,Hannuksela:2020xor,Robertson:2020mfh,Ryczanowski:2020mlt,Yu:2020agu}. 
This identification could take place by matching \gw and electromagnetic image properties when four \gw images are available~\citep{Hannuksela:2020xor}.
With two images, the number of hosts could also be constrained~\citep{Sereno:2011ty,Yu:2020agu}, but to a lesser degree due to degeneracies with the lens and source alignment and uncertainties introduced by micro/millilensing -- although strong lensing by galaxy clusters might allow us to identify a single cluster candidate~\citep{Smith:2017mqu,Smith:2018gle,Smith:2019dis,Robertson:2020mfh,Ryczanowski:2020mlt}.
Moreover, strong lensing could have produced additional images below the noise threshold.
We perform a further investigation of such sub-threshold counterparts in the next section.

    \subsection{Search for sub-threshold lensed images}
    \label{sec:subthreshold}
    Here we search for sub-threshold counterpart images of the O3a events from GWTC-2
that would not have been identified as confident detections by the search pipelines used in~\citet{GWTC2}.
As lensed images could in principle appear anywhere in the entire O3a data,
we perform targeted template bank searches for these sub-threshold lensed counterparts over the whole O3a strain data set, following the data selection criteria described in~\citet{GWTC2}.
We employ two matched-filter searches based on
the \gstlal~\citep{Cannon:2011vi, Messick:2016aqy, Hanna:2019ezx, Sachdev:2019vvd}
and \pycbc~\citep{Usman:2015kfa, Nitz:2018rgo, Nitz:2019pycbc, Davies:2020tsx}
pipelines,
adapted to the lensing case in similar ways as in~\citet{Li:2019osa} and~\citet{McIsaac:2019use}.

\subsubsection{Search methods and setups}
The lensed hypothesis states that the intrinsic masses and spins will remain consistent between multiple lensed images of the same event.
Hence, we can perform searches that specifically target sub-threshold lensed counterparts of known events
by creating reduced banks of template waveforms with masses and spins close to those inferred for the primary event.
We use the public posterior mass and spin samples released with GWTC-2~\citep{gwosc:gwtc2} to create these targeted template banks.
This ensures that the known events will match well with the templates while simultaneously decreasing the \far of the search for similar events,
potentially returning new candidates that did not reach the search threshold in~\citet{GWTC2}. 
\gstlal{'s} reduced banks contain between $173$ and $2698$ templates per search,
while for each \pycbc search we select a single aligned-spin template.
The construction of these template banks closely follows~\citet{Li:2019osa,McIsaac:2019use} and is further detailed in Appendix~\ref{app:subthr}.
Template waveforms are generated using the aligned-spin SEOBNRv4\_ROM waveform~\citep{Bohe:2016gbl, Purrer:2014fza, Purrer:2015tud}
for both pipelines and all events,
with the exception of GW190425 in the \pycbc search, where we use the TaylorF2 model~\citep{Blanchet:1995ez,SathyaDhurandhar:1991,Poisson:1997ha,DamourJaranowskiSchaefer:2000,Mikoczi:2005dn,Blanchet:2005tk,Arun:2008kb,Buonanno:2009zt,Faye:2012we,Bohe:2013cla,Blanchet:2013haa,Bohe:2015ana,Mishra:2016whh}.

Given these template banks, each search pipeline proceeds with configurations and procedures as outlined in~\citet{GWTC2}
to produce a priority list of potential lensed candidates matching each target event.
To rank these, each pipeline uses a different method to estimate \far[s].

\gstlal first identifies matched-filter triggers from one or more of the Hanford, Livingston, and Virgo data streams.
Coincidences are identified with the same settings as in~\citet{GWTC2}.
From each candidate's recovered parameters, a likelihood-ratio ranking statistic is computed~\citep{Sachdev:2019vvd}.
Single-detector triggers are penalized using machine-learning based predictions \citep[iDQ;][]{Essick:2020qpo,Godwin:2020weu}
whereas for coincident triggers, no data quality products are used.
We estimate the \far of a trigger by comparing with the distribution of the ranking statistic
from all non-coincident noise triggers, used to characterize the noise distribution,
over the O3a data set.

\pycbc also first identifies single-detector matched-filter triggers, with a reduced clustering window compared to the GWTC-2 configuration (from 1\,s to 0.01\,s).
These are tested for time coincidence between detectors and are required to have an \mbox{SNR $\geq 4$} in at least two detectors.
While in the GWTC-2 analysis the \pycbc search was limited to the Hanford and Livingston detectors,
here we also include Virgo data, using the methods described in~\citet{Davies:2020tsx} to analyse the three detector network.
\far[s] are estimated from a noise background measured using time-shifted data.
All triggers within 0.1\,s of the times of the events in GWTC-2 are removed from both
the foreground (observed coincident events)
and the background.

Candidates from both pipelines are further vetted by a sky localization consistency test against the targeted GWTC-2 event,
as lensed images of the same event should come from consistent sky locations
but the matched-filter searches do not check for this.
For each new candidate, we generate a sky localization map $p(\Omega)$ using \bayestar \citep{Singer:2015ema},
with $\Omega$ denoting parameters that specify the sky location.
We compute the percentage overlap $O_{\rm 90\% CR}$ of the 90\% credible regions
between the sky localization $q(\Omega)$ of a GWTC-2 event and the sky localization $p(\Omega)$ of a sub-threshold event candidate as
\begin{equation}
O_{\rm 90\% CR} = \frac{\mathbf{1}_{\rm 90\% CR} \left[ p(\Omega)q(\Omega) \right] \mathrm{d}\Omega}{{\rm min} (\mathbf{1}_{\rm 90\% CR} \left[ p(\Omega) \right] \mathrm{d}\Omega , \mathbf{1}_{\rm 90\% CR} \left[ q(\Omega) \right] \mathrm{d}\Omega)},
\end{equation}
where $\mathbf{1}$ is the indicator function.
To avoid false dismissal at this step, we only veto candidates with \mbox{$O_{\rm 90\% CR}=0$}.
All candidates with non-vanishing localization overlap are kept for further follow-up with
data quality checks as discussed in Sec.~\ref{sec:events}
and with the joint-\pe methods described in Sec.~\ref{sec:jointpe}.

\subsubsection{Results}
In Table~\ref{table:subthresh-search-results},
we list the eight candidates with \far $<1$ in $16$ years from the \emph{individual} targeted searches for counterparts of the $39$ detections reported in GWTC-2 found by at least one pipeline.
Six of these are unique candidates.
This number, compared with $\sim2$ expected noise events above this \far from the number of searches performed,
is consistent with additional astrophysical signals being present in the data set.
However, in this work, we do not assess in detail the probability of astrophysical origin for each of these.
The reported \far[s] also do not indicate how likely each trigger is to be a lensed counterpart of the targeted event,
but only how likely it is to obtain a trigger with a similar ranking statistic from a pure noise background
using these reduced template banks.
Three of these candidates were also recovered
with high probability of an astrophysical origin
in the 3-OGC open-data search~\citep{Nitz:2021uxj}, which used a broad template bank.
Five of them are also included with $p_\mathrm{astro}>0.5$ in the extended catalog GWTC-2.1~\citep{LIGOScientific:2021usb}.
Candidates matching one or both of these catalogs are marked with footnotes in Table~\ref{table:subthresh-search-results}.

\begin{figure}
\includegraphics[width=\columnwidth]{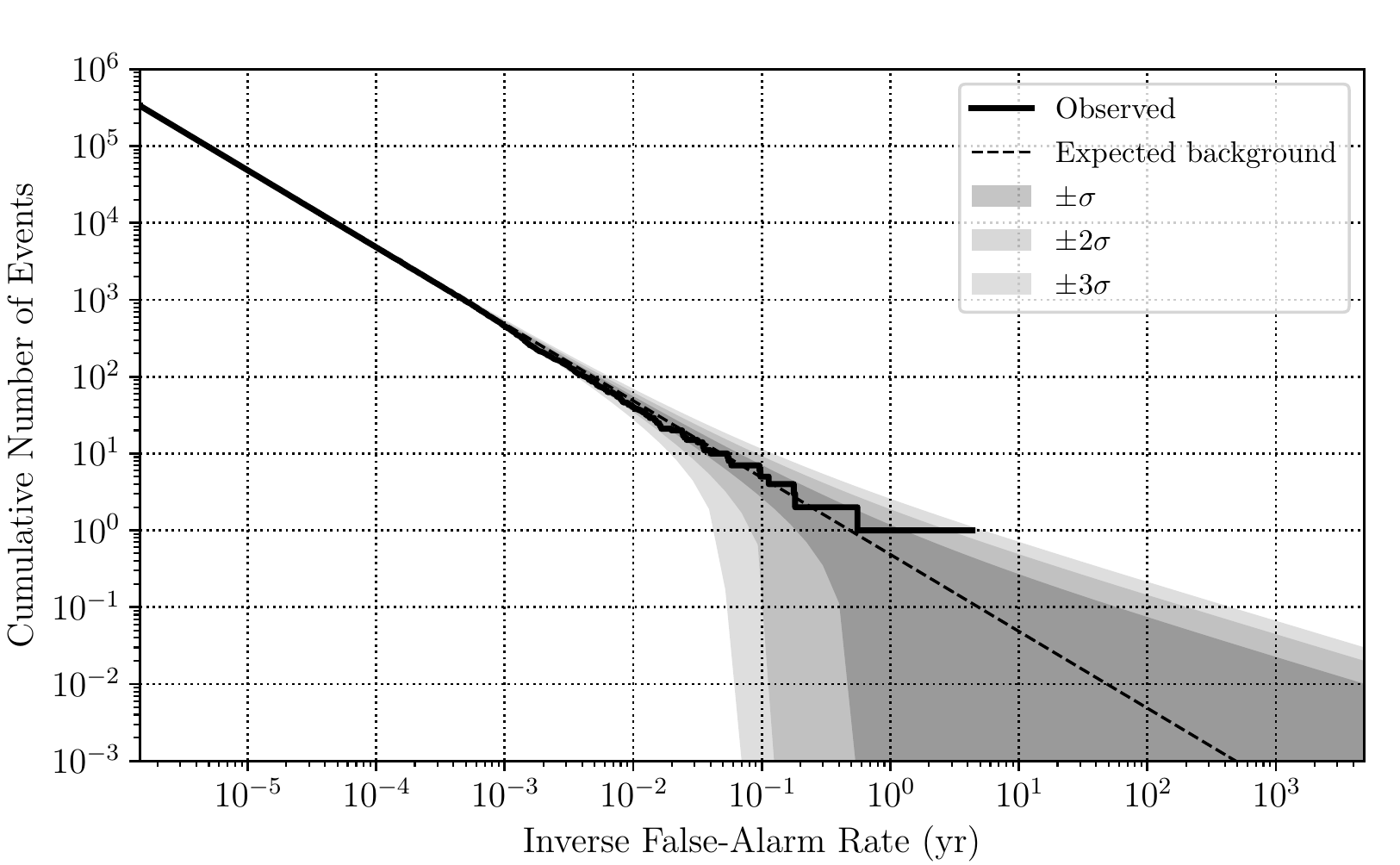}\\
\includegraphics[width=\columnwidth]{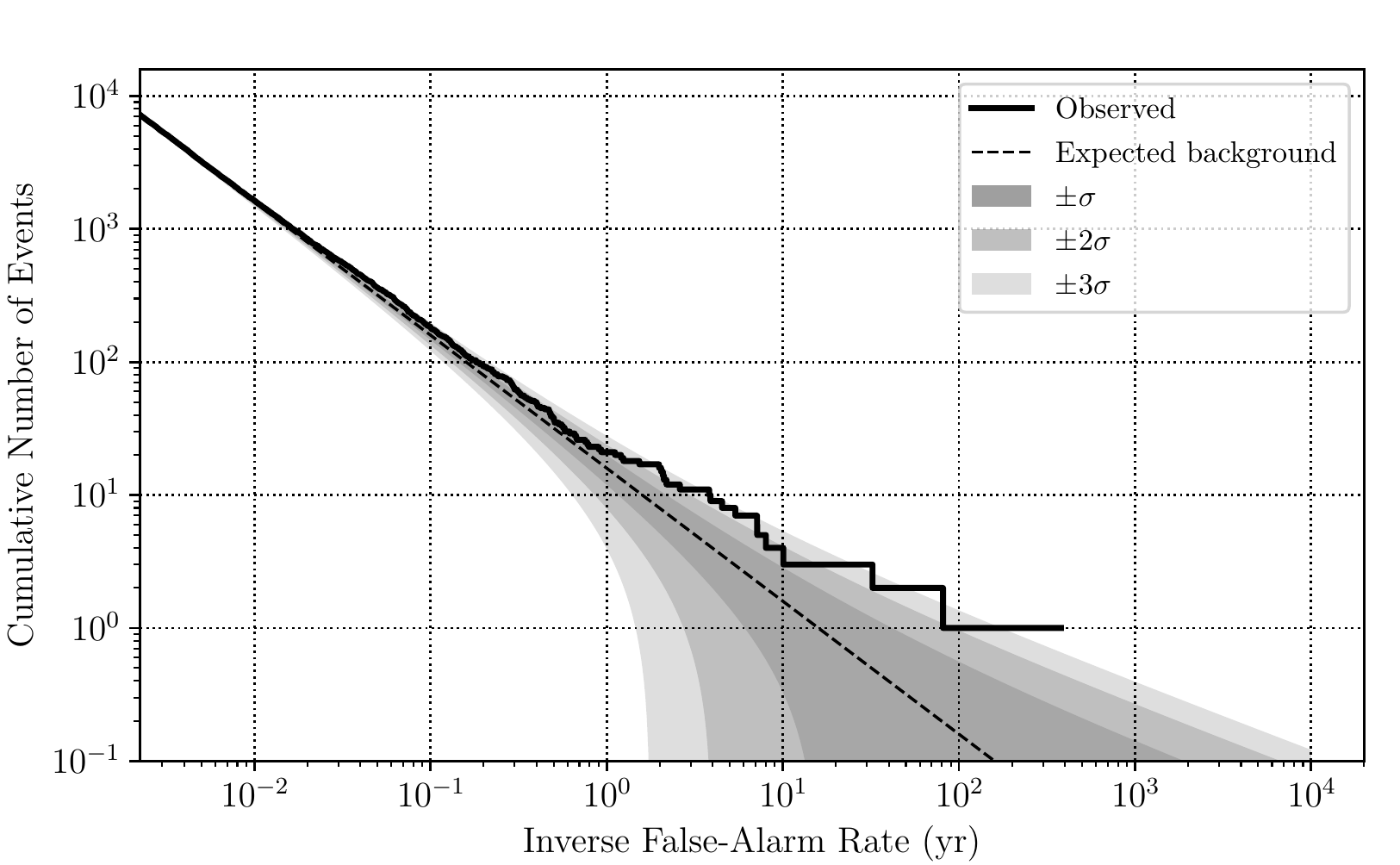}
\caption{\label{fig:subthresh-plots}
Combined results from the \NUMEVENTS{} sub-threshold searches
with the \gstlal pipeline (top panel)
and \pycbc pipeline (bottom panel).
Each panel shows,
as a solid line,
the cumulative number of coincident triggers
(observed)
with inverse \far[s] greater than or equal to a given value.
The dashed line is the expected distribution of background triggers,
with the gray bands indicating uncertainties
in multiples of the standard deviation $\sigma$ of a Poisson distribution.
For \gstlal, the results for this plot
are obtained by a search over all O3a data
using a combined bank from the \NUMEVENTS{} targeted banks.
For \pycbc, the \far[s] are from the individual searches,
but for triggers found by several of the single-template searches,
their inverse \far[s] have been summed.
}
\end{figure}

In contrast, Fig.~\ref{fig:subthresh-plots} shows the \emph{combined} search results from all $39$ targets for each pipeline: \gstlal (top panel) and \pycbc (bottom panel), excluding triggers that correspond to other detections already reported in GWTC-2. Each panel shows the cumulative number of coincident triggers (observed) with inverse \far[s] greater than or equal to a given threshold value. For \gstlal, the combined results are obtained by a search over all O3a data using a combined template bank from the $39$ targeted banks. For \pycbc, the \far[s] are obtained from the individual searches, but for triggers being found in several single-template searches, their inverse \far[s] are summed. In the same figure, we compare these results with estimated background distributions, accounting for the fact that we have re-analyzed the same data set of $\sim$\,150~days multiple times, and find a slight excess in the rate of foreground triggers at high inverse \far[s].

\begin{table*}[t]
\caption{\label{table:subthresh-search-results}
Candidates from individual sub-threshold searches for strongly-lensed counterpart images of the \NUMEVENTS{} O3a events from GWTC-2.
}
\hspace{-1.75cm}
\resizebox{1.1\textwidth}{!}{
\begin{tabular}{ccrrllrcccc}
\tableline \tableline
UTC time & GWTC-2 targeted event & $|\Delta t |$ [d] & $(1+z)\Mc$  & \multicolumn{2}{c}{\far $\left[\text{yr}^{-1}\right]$} & $O_{\rm 90\% CR}$ [\%] & $\log_{10}\cohratio$ (\linf) \\
 & & & [$\Msun$] & \pycbc & \gstlal & & ($\Delta\phi$: $0$, $\pi/2$, $\pi$, $3\pi/2$) \\
\tableline
2019 Sep 25 23:28:45\tablenotemark{a}\tablenotemark{b} & GW190828\_065509 & $28.69$ & $17.3$ & $0.003$ & $98.681$ & $0.0\%$ & $-$ \\
2019 Apr 26 19:06:42\tablenotemark{b} & GW190424\_180648 & $2.04$ & $65.5$ & $-$ & $0.017$ & $63.8\%$ & $(-5.8, -5.8, -5.9, -5.6)$ \\
2019 Jul 11 03:07:56 & GW190421\_213856 & $80.23$ & $47.7$ & $0.032$ & $0.341$ & $1.2\%$ & $(+2.3, +1.1, +1.1, +2.6)$ \\
2019 Jul 25 17:47:28\tablenotemark{a}\tablenotemark{b} & GW190728\_064510 & $2.54$ & $9.0$ & $-$ & $0.038$ & $0.0\%$ & $-$ \\
2019 Jul 11 03:07:56 & GW190731\_140936 & $20.46$ & $47.4$ & $0.045$ & $0.944$ & $2.9\%$ & $(+2.6, -1.2, -1.6, +0.9)$ \\
2019 Aug 05 21:11:37\tablenotemark{b} & GW190424\_180648 & $103.13$ & $68.8$ & $-$ & $0.051$ & $26.9\%$ & $(-1.1, +0.6, -0.3, -0.7)$ \\
2019 Jul 11 03:07:56 & GW190909\_114149 & $60.36$ & $49.0$ & $0.053$ & $1.196$ & $12.6\%$ & $(+3.5, +2.2, +3.4, +2.9)$ \\
2019 Sep 16 20:06:58\tablenotemark{a}\tablenotemark{b} & GW190620\_030421 & $88.71$ & $53.3$ & $0.055$ & $1.389$ & $49.5\%$ & $(+1.7, +3.6, +2.1, -3.2)$ \\

\tableline
\end{tabular}
} 
\tablecomments{
The first column shows the UTC time of the newly found sub-threshold candidate.
The second column lists the targeted O3a event from the catalog GWTC-2;
see Table~IV and Table~VI of \citet{GWTC2} for details of these.
The third column shows the absolute time difference between the candidate and the targeted event.
The fourth column shows the redshifted chirp mass of the template that generated the trigger.
The fifth and sixth columns show the corresponding \far[s] from the individual search for the target from the second column, from each of the two search pipelines
(\gstlal and \pycbc),
if the candidate has been recovered by it.
The seventh column shows the percentage overlap of the $90\%$ sky localization regions
between the candidate and the targeted event,
from the pipeline with the lower \far.
The eighth column shows the coherence ratio $\cohratio$ for the pair from the \linf joint-\pe follow-up
(only for candidate pairs with a localization overlap $>$ $0\%$).
Candidates are only reported here if they pass a \far threshold of $<$ $1$ in $16$~years in at least one pipeline,
and are sorted in ascending order by the lowest \far from either pipeline.
If the same new trigger was found with sufficient \far by more than one search for different targets, all occurrences are included, and the \pe follow-up is conducted separately for each pair.
Candidates that have since also been reported by other searches are marked with footnotes.
}
\tablenotetext{a}{also included in 3-OGC~\citep{Nitz:2021uxj}}
\tablenotetext{b}{also included in GWTC-2.1~\citep{LIGOScientific:2021usb}}
\end{table*}

Instead, we perform follow-up analyses of the lensing hypothesis under the assumption of astrophysical origin,
aiming to determine for each candidate pair in Table~\ref{table:subthresh-search-results}
whether it is more consistent with a pair of images of a single lensed event or with two independent astrophysical events.
After taking into account the initial \far thresholds, sky localization overlap, and data-quality checks,
we have followed up six candidate pairs through \linf joint Bayesian \pe
as described in Sec.~\ref{sec:jointpe-linf}.
No special mitigation steps were required for data-quality reasons on any of the new candidates.
The results are included in Table~\ref{table:subthresh-search-results}.

Compared with the results for GWTC-2 pairs in Table~\ref{table:pe-blu},
the \linf coherence ratios alone are insufficient to provide evidence of lensing
while keeping in mind selection effects and prior odds.
As another cross-check, we have also analysed the pair with the highest \linf coherence ratio $\cohratio$
(the candidate on 2019 September 16 found by the GW190620\_030421 \pycbc search)
with the \hanabi pipeline described in Sec.~\ref{sec:jointpe-hanabi}.
As with all pairs previously tested (see Table~\ref{table:pe-blu}),
after the inclusion of population priors and selection effects, there is no evidence favoring the lensing hypothesis for this pair either,
with population-weighted coherence ratio $\log_{10}(\left.\cohratio\right|_\mathrm{pop})=-0.1$
and Bayes factor $\log_{10}(\Blu)=-3.2$.

As lensing can produce more than two images of the same source,
cases where several searches find the same trigger are of particular interest.
We find that the same candidate on 2019 July 11 has been found with low \far[s] by three searches
(targeting the GWTC-2 events  GW190421\_213856, GW190731\_140936, and GW190909\_114149).
In addition, the trigger on 2019 August 05 is only found with sufficient \far for inclusion in Table~\ref{table:subthresh-search-results}
by a single \gstlal search (for GW190424\_180648),
but was also recovered by those for GW190413\_052954 and GW190803\_022701 with \far[s] just below the cut.
However, the GWTC-2 pairs involved in these possible quadruple sets
have already been significantly disfavored by the \hanabi analysis
including population priors and selection effects.
We also expect such multiple matches from an unlensed \bbh population
due to the clustering of the GWTC-2 events in parameter space~\citep{GWTC2,GWTC2:rates}.

Also, as discussed in detail in \citet{McIsaac:2019use},
if any high-mass \gw detections are interpreted as highly magnified images of lower-mass sources,
then counterpart images for these would be more likely.
However, we did not find any promising sub-threshold candidates for the five events
discussed under the lensing magnification hypothesis in Sec.~\ref{sec:individualevents}.

In summary, the sub-threshold searches can recover additional promising candidates that were not included in GWTC-2,
which match other events closely and, in that sense, are consistent with the lensing hypothesis.
However, we do not find sufficient evidence that they are indeed lensed images,
as the set of results is also consistent with a population of physically independent and only coincidentally similar events.

\section{Search for microlensing effects} 
\label{sec:microlensing}
\textit{Microlensing} by smaller lenses produces image separations on the order of microarcseconds.
For \gw[s], it can also induce frequency-dependent wave-optics effects similar to femtolensing of light~\citep{Nakamura:1997sw,Takahashi:2003ix}.
More specifically, when the characteristic wavelengths are comparable to the Schwarzschild radius of the lens, i.e.,
$\lambda_\mathrm{GW}\sim R^\mathrm{lens}_\mathrm{Sch}$,
it causes frequency-dependent magnification of the waveform.
Moreover, the characteristic lensing time-delay due to microlensed images can be shorter than the \gw signal duration, causing potentially observable beating patterns on the waveform~\citep{Cao:2014oaa,Jung:2017flg,Lai:2018rto,Christian:2018vsi,Dai:2018enj,Diego:2019lcd,Diego:2019rzc,Pagano:2020rwj,Cheung:2020okf,Mishra:2021abc}, due to waveform superposition.
To observe \gw microlensing, we search for these beating patterns instead of the time-dependent change in the flux traditionally observed for microlensing in electromagnetic signals.

Here we search for microlensing by isolated point masses. 
The microlensed waveform has the form
\begin{equation}
h^\mathrm{ML}(f;\theta_\mathrm{ML}) = h^\mathrm{U}(f;\theta) \, F(f; \MzL, y)\,,
\label{eq:wl_lens_hyp}
\end{equation}
where $h^\mathrm{ML}$ and $h^\mathrm{U}$ are the microlensed and unlensed waveforms in the frequency domain, respectively.
$\theta$ represents the set of parameters defining an unlensed \gw signal,
while \mbox{$\theta_\mathrm{ML} = \{\theta, \MzL, y\}$}.
$F(f; \MzL, y)$ is the frequency-dependent lensing magnification factor,
which is a function of the redshifted lens mass \mbox{$\MzL = M_\mathrm{L} (1+\zL)$} and dimensionless impact parameter $y$, 
given in Eq. 2 of~\citep{Lai:2018rto}.
The search involves re-estimating the parameters of previously identified events under the microlensed hypothesis as defined in Eq.~\eqref{eq:wl_lens_hyp},
including those of the potential lens.

To measure the evidence of lensing signatures in a signal, we define a Bayes factor
$\BMLU$, which is the evidence ratio between the microlensed and unlensed hypotheses.
Higher positive values correspond to support for lensing. 
~\cite{Hannuksela:2019kle} searched for similar beating patterns due to point mass lenses in the O1 and O2 data,
using an upper lens mass prior cutoff $\MzL \lesssim 10^5 \Msun$.
They reported no evidence for such lensing patterns above $\log_{10}\BMLU>0.2$.

For O3a, we analyze the 36 events from~\cite{GWTC2} that confidently have both component masses above $3\Msun$
and search for microlensing signatures following the same method as in~\cite{Hannuksela:2019kle}.
We perform \pe using \bilby~\citep{Ashton:2018jfp, Romero-Shaw:2020owr}
and the nested sampling algorithm dynesty~\citep{10.1093/mnras/staa278}.
For each event, we perform two \pe runs using both unlensed and microlensed templates.
For the unlensed case, which is similar to the usual \pe analysis, equivalent prior settings and data dictionaries such as strain data and power spectral densities (PSDs) are used as in~\citet{GWTC2}.
The analysis uses the \IMRXPHM~\citep{Pratten:2020ceb} waveform for most events,
except for GW190521, which is analyzed using the \NRSur waveform~\citep{Varma:2019csw}
and for the least massive event GW190924\_021846 where the \IMRP waveform is used.
The prior on $\MzL$ is log uniform in the range $[1$--$10^5~\Msun]$, above which the effect of microlensing is relatively small for the LIGO--Virgo sensitivity band.
The impact parameter prior is $p(y)\propto y$ between $[0.1,3]$, chosen due to geometry and isotropy~\citep{Lai:2018rto}.

\begin{figure}
\includegraphics
  [width=1.0\hsize]
  {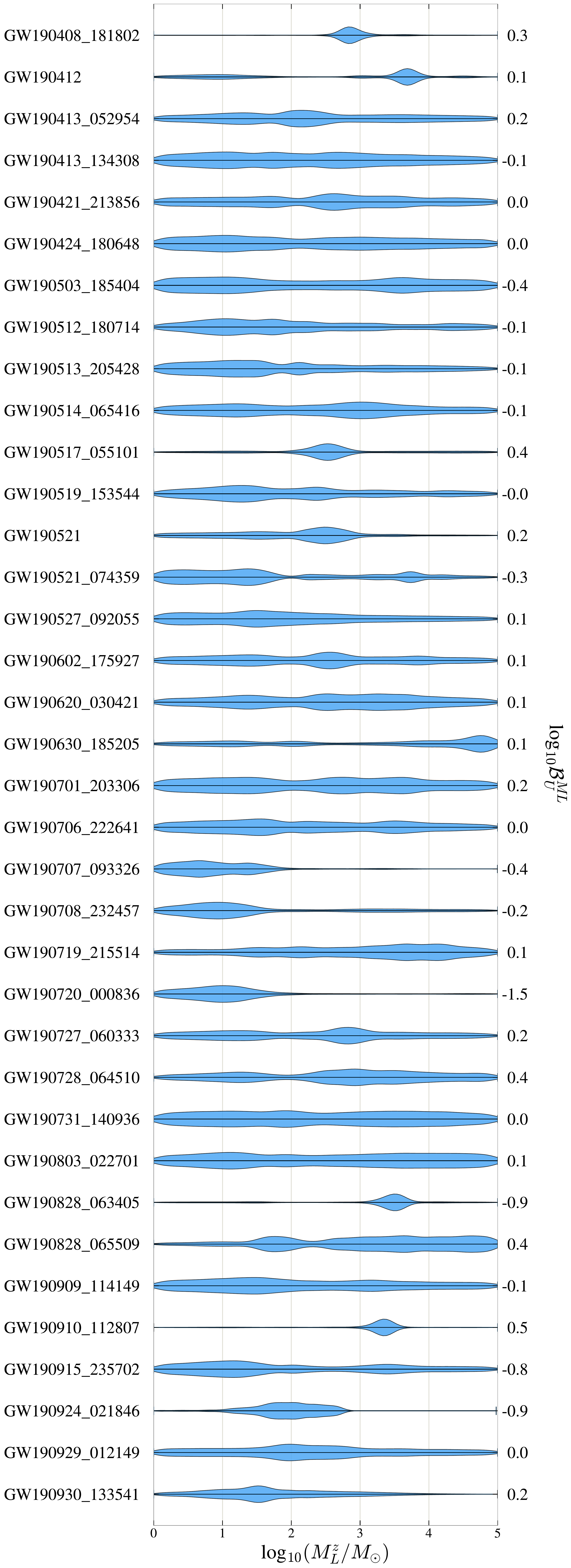}
\caption{The marginalized posterior distribution of redshifted lens mass $\MzL$ and $\log_{10}\BMLU$ between microlensed and unlensed hypotheses.
The corresponding $\log_{10}$ Bayes factors are noted to the right of the plot. 
We find no evidence of microlensing by point mass lenses. 
}
\label{fig:violin_plot_o3a}
\end{figure}

In Fig.~\ref{fig:violin_plot_o3a} we show violin plots of marginalized posterior distributions for the redshifted lens mass for each event,
as well as the Bayes factors between the microlensed and unlensed hypotheses.
The broad $\MzL$ posteriors correspond to broad posteriors on the impact parameter $y$, which is not well constrained for unlensed cases.
In terms of Bayes factors, there is no substantial evidence of microlensing with a maximum $\log_{10}\BMLU = 0.5$ for the event GW190910\_112807.
Additionally, as can be seen in Appendix~\ref{app:injection_study_microlensing}, statistical fluctuations of the $\log_{10}$ Bayes factors for injections without microlensing can be as high as 0.75. 
Thus, the observed Bayes factors are already by themselves consistent with random noise fluctuations
and do not significantly favor the microlensing hypothesis for any of the events. 
The resulting posterior odds $\mathcal{O}_\mathrm{U}^\mathrm{ML}$,
which are the products of Bayes factors and the low prior odds of microlensing~\citep{Lai:2018rto},
would be even lower.
Thus, we find no evidence of microlensing in this study.

We searched for microlensing due to isolated point masses.
More complex models in which point mass lenses embedded in an external macromodel potential such as galaxies and galaxy clusters~\citep{Diego:2019lcd,Cheung:2020okf,Mishra:2021abc} 
can produce additional modulation on the magnified waveform, which could also prove important in the LIGO--Virgo frequency band.
Future searches could be extended to cover a broader range of microlensing models.

\section{Conclusions and outlook}
\label{sec:conclusion}
We have searched for gravitational lensing effects on the \gw observations from O3a,
the first half of the third LIGO--Virgo observing run, finding no strong evidence of lensing.
First, we outlined estimates for the rate of strongly lensed \gw[s].
Second, presuming a non-observation of lensing, we constrained the \bbh merger-rate density at high redshift.
Third, we used merger-rate density models obtained through the non-observation of a \sgwb to estimate the \gw lensing rate. 

Next, we performed an analysis of apparent high-mass events under the hypothesis that they are lensed signals from lower-mass sources,
finding that the highest-mass \bbh[s] from O3a could be consistent with component masses below the PISN mass gap, 
while GW190425 and GW190426\_152155 would require extreme magnifications to be compatible with the Galactic \bns population.
This hypothesis is at the moment mainly disfavored by the expected lensing rates,
but in the future, more quantitative constraints could also be set by connecting these magnification results with lens modeling to make predictions for the appearance of multiple images or the possibility of microlensing.

We then searched for signatures of multiple lensed images from a single source through several methods.
We first investigated the parameter consistency among all pairs of O3a events from GWTC-2 using a posterior-overlap method,
finding no significant event pairs but identifying several interesting candidates with high overlap.

We followed up on these candidate pairs using two detailed joint-\pe analyses,
finding high parameter consistency for 11 pairs.
However, after the inclusion of a more appropriate population prior, selection effects, and the prior odds against the lensing hypothesis, these candidates do not provide sufficient evidence for a strong lensing claim.

Moreover, we used two targeted matched-filter approaches to search for additional lensed images of the known events that could be hidden beneath the thresholds of the corresponding broader analyses used to produce GWTC-2, identifying six new candidates.
After follow-up by joint \pe,
we found no evidence to conclude that any of these sub-threshold triggers are lensed images.

Finally, we analyzed 36 events from GWTC-2 for microlensing effects
by performing full \pe with waveforms incorporating microlensing by point mass lenses. 
We found no evidence of microlensing. 

In summary, our results on O3a data are consistent with the expected low rate of lensing at current detector sensitivities.
However, improved analysis methods and lens modeling may allow digging deeper into potential lensing effects. 
Electromagnetic follow-up of lensing candidates, even if they are not significant enough based on the \gw data alone,
could also be promising~\citep{Sereno:2011ty, Smith:2017mqu,Hannuksela:2020xor, Yu:2020agu}. 
With the current generation of detectors further improving their sensitivity and the global network being extended~\citep{Aasi:2013wya},
the chances of detecting clear lensing signatures will improve,
and the field will offer many possibilities at the latest with
third-generation~\citep{0264-9381-27-8-084007,Evans:2016mbw,Reitze:2019iox,Maggiore:2019uih} and space-based detectors~\citep{Audley:2017drz,Hu:2017mde}
and their expected cosmological reach.

\acknowledgments
{
This material is based upon work supported by NSF’s LIGO Laboratory which is a major facility
fully funded by the National Science Foundation.
The authors also gratefully acknowledge the support of
the Science and Technology Facilities Council (STFC) of the
United Kingdom, the Max-Planck-Society (MPS), and the State of
Niedersachsen/Germany for support of the construction of Advanced LIGO 
and construction and operation of the GEO600 detector. 
Additional support for Advanced LIGO was provided by the Australian Research Council.
The authors gratefully acknowledge the Italian Istituto Nazionale di Fisica Nucleare (INFN),  
the French Centre National de la Recherche Scientifique (CNRS) and
the Netherlands Organization for Scientific Research, 
for the construction and operation of the Virgo detector
and the creation and support  of the EGO consortium. 
The authors also gratefully acknowledge research support from these agencies as well as by 
the Council of Scientific and Industrial Research of India, 
the Department of Science and Technology, India,
the Science \& Engineering Research Board (SERB), India,
the Ministry of Human Resource Development, India,
the Spanish Agencia Estatal de Investigaci\'on,
the Vicepresid\`encia i Conselleria d'Innovaci\'o, Recerca i Turisme and the Conselleria d'Educaci\'o i Universitat del Govern de les Illes Balears,
the Conselleria d'Innovaci\'o, Universitats, Ci\`encia i Societat Digital de la Generalitat Valenciana and
the CERCA Programme Generalitat de Catalunya, Spain,
the National Science Centre of Poland and the Foundation for Polish Science (FNP),
the Swiss National Science Foundation (SNSF),
the Russian Foundation for Basic Research, 
the Russian Science Foundation,
the European Commission,
the European Regional Development Funds (ERDF),
the Royal Society, 
the Scottish Funding Council, 
the Scottish Universities Physics Alliance, 
the Hungarian Scientific Research Fund (OTKA),
the French Lyon Institute of Origins (LIO),
the Belgian Fonds de la Recherche Scientifique (FRS-FNRS), 
Actions de Recherche Concertées (ARC) and
Fonds Wetenschappelijk Onderzoek – Vlaanderen (FWO), Belgium,
the Paris \^{I}le-de-France Region, 
the National Research, Development and Innovation Office Hungary (NKFIH), 
the National Research Foundation of Korea,
the Natural Science and Engineering Research Council Canada,
Canadian Foundation for Innovation (CFI),
the Brazilian Ministry of Science, Technology, and Innovations,
the International Center for Theoretical Physics South American Institute for Fundamental Research (ICTP-SAIFR), 
the Research Grants Council of Hong Kong,
the National Natural Science Foundation of China (NSFC),
the Leverhulme Trust, 
the Research Corporation, 
the Ministry of Science and Technology (MOST), Taiwan,
the United States Department of Energy,
the Kavli Foundation,
and the Gordon and Betty Moore Foundation.
The authors gratefully acknowledge the support of the NSF, STFC, INFN and CNRS for provision of computational resources.

{\it We would like to thank all of the essential workers who put their health at risk during the COVID-19 pandemic, without whom we would not have been able to complete this work.}

\software{
Analyses in this paper made use of 
\lal~\citep{lalsuite},
the
\gstlal~\citep{Cannon:2011vi, Messick:2016aqy, Hanna:2019ezx, Sachdev:2019vvd}
and \pycbc~\citep{Usman:2015kfa, Nitz:2018rgo, Nitz:2019pycbc, Davies:2020tsx} pipelines;
Bayesian inference with
\textsc{CPNest} \citep{CPNest},
\textsc{PyMultinest} \citep{Feroz:2009,Feroz:2013},
\bilby~\citep{Ashton:2018jfp, Smith:2019ucc,Romero-Shaw:2020owr},
and \linf~\citep{Veitch:2014wba};
as well as the packages
\textsc{NumPy} \citep{Harris:2020xlr},
\textsc{SciPy} \citep{2020SciPy-NMeth},
\textsc{Astropy} \citep{astropy:2013,astropy:2018},
\textsc{IPython} \citep{ipython}, 
and \textsc{ligo.skymap} \citep{ligoskymap}.
Plots were produced with
\textsc{Matplotlib} \citep{matplotlib},
and \textsc{Seaborn} \citep{seaborn}.
}

}

\vspace{5mm}

\appendix

\section{Lensing statistics supplementary}
\label{app:lens_statistics_supplementary}
Assuming a specific \bbh formation channel, we can estimate the lensing rate for merger signals from that population. 
For example, suppose \bbh[s] form as a consequence of isolated binary evolution. 
In that case, one can theoretically model \bbh formation assuming that it traces the star-formation rate, modulated by the delay time distribution and by the stellar metallicity evolution~\citep{Belczynski:2005mr,Belczynski:2010tb,Dominik:2013tma,Marchant:2018kun,Eldridge:2018nop,Neijssel:2019irh,Boco:2019teq,Santoliquido:2020axb}.
However, note that if the \bbh[s] form through other means or through multiple channels, the merger-rate density could be different~\citep[e.g.,][]{Miller:2008yw,Antonini:2016gqe,Rodriguez:2018rmd,Fragione:2020nib,DeLuca:2020bjf,Antonini:2020xnd,Wong:2020ise,Zevin:2020gbd,Bouffanais:2021wcr}.

Here we assume two models for the merger-rate density. 
We base the first model on the assumption that the merger-rate density of the observed \bbh[s] traces the star-formation rate density and the \bbh[s] originate from Population I/II stars. 

In this work, we did not consider the contribution of Population III stars.
Population III stars have not been observed yet, and their physical properties, binary fraction, and initial mass function are still a matter of debate~\citep{Nakamura:2000ez, Madau:2001sc, Bromm:2001bi, Schaerer:2001jc, Norman:2008px, Machida:2008ni, 2018ApJ...857...46I}.
As such, the contribution of Population III \bbh[s] to gravitational-wave sources is
also uncertain~\citep[e.g.,][]{1984MNRAS.207..585B, 2012A&A...541A.120K, Belczynski:2016ieo,Liu:2020ufc}.
Should Population III stars dominated the \bbh formation at high redshift,
our results would need to be re-interpreted.

The first model, which we label Model A, uses the following fits that bracket the available population synthesis results from the literature \citep[e.g.,][]{Belczynski:2005mr,Belczynski:2010tb,Dominik:2013tma,Marchant:2018kun,Eldridge:2018nop,Neijssel:2019irh,Boco:2019teq,Santoliquido:2020axb}:
\begin{equation} \label{app:eq:rate_rp}
\begin{split}
 \mathcal{R}_\textrm{m}^{\rm min}(z_\textrm{m}) &= \frac{a_1~e^{a_2 z_\textrm{m}}}{a_3+ e^{a_4 z_\textrm{m}}} \rm Gpc^{-3}\,yr^{-1} \,, \\
 \mathcal{R}_\textrm{m}^{\rm max}(z_\textrm{m}) &= \frac{b_1~e^{b_2 z_\textrm{m}}}{b_3 + e^{b_4 z_\textrm{m}}} \rm Gpc^{-3}\,yr^{-1} \,,
\end{split}
\end{equation}
where the fitting parameters $a_1=58.497$, $a_2=2.06424$, $a_3=2.82338$, $a_4=2.52898$, $b_1=105356$, $b_2=1.30278$, $b_3=2714.36$, and $b_4=2.22903$.

We base the second model, Model B, on the assumption that the merger-rate density follows the \citet{Madau:2014bja} ansatz:
\begin{equation} \label{app:eq:rate_rp_stochastic}
    \mathcal{R}_{\rm m}(z_\textrm{m}; R_0, \alpha) = R_0 \frac{(1+z_\textrm{m})^\kappa}{1+[(1+z_\textrm{m})/(1+z_p)]^{(\gamma+\kappa)}}\,.
\end{equation}

To constrain the merger-rate density at high redshift, we assume that no strong lensing has occurred (Sec.~\ref{ssec:implications_of_non_detection}).
We further assume that events occur following a Poisson process.

Let us now assume Model B for the merger-rate density, Eq.~\ref{app:eq:rate_rp_stochastic}.
The distribution of merger-rate density parameters, given that no strong lensing has occurred, 
\begin{equation}
    p(R_0, \kappa, \gamma, z_p|N, \{d_i\}) \propto \mathcal{W} \times p(R_0, \kappa, \gamma, z_p|\{d_i\})\,,
\end{equation}
where $p(R_0, \kappa, \gamma, z_p|\{d_i\})$ follows the posterior distribution of parameters inferred from LIGO--Virgo population studies~\citep{GWTC2:rates}, and
\begin{equation}
    \mathcal{W} = \frac{N_{\rm avg}(R_0, \kappa, \gamma, z_p)^N \exp[{- N_{\rm avg}(R_0, \kappa, \gamma, z_p)}] }{N!}\,,
\end{equation}
with $N$ being the number of observed, strongly lensed \gw signals,
and $N_{\rm avg}(R_0, \kappa, \gamma, z_p)$ the expected number of events within a time $\Delta t$. 
Here, like in Sec.~\ref{sec:statistics}, we do not account for detector downtime, and instead as a proxy presume that the detectors are always online.
The $R_0$ and $\kappa$ value is measured at a low redshift \citep{GWTC2:rates}. 
The $\gamma$ and $z_p$ values are unconstrained here and thus match an uninformative prior, with $p(\gamma)=\mathcal{S}\mathcal{N}(5,10,3)$ being a split normal distribution and $p(z_p)$ being uniformly distributed between $[0,4]$. 
The above equations give all the necessary ingredients to forecast the rate of strongly lensed events 
and place constraints on the merger-rate density 
based on the number of lensed signals observed by LIGO and Virgo.

\section{Construction of sub-threshold counterpart search template banks}
\label{app:subthr}
For the \gstlal and \pycbc searches for sub-threshold lensed counterparts (Sec.~\ref{sec:subthreshold})
the targeted template banks for each event are constructed starting from a certain choice of posterior distributions released with GWTC-2~\citep{GWTC2,gwosc:gwtc2},
aiming for a reduced-size template bank that is effective at recovering signals similar to the primary observed event.

For the \gstlal pipeline, we start, for all but three of the O3a events from GWTC-2,
from non-spinning posteriors obtained with the \IMRD waveform~\citep{Husa:2015iqa, Khan:2015jqa}.
In three cases, we instead start from posteriors obtained with the \IMRP waveform~\citep{Hannam:2013oca,Bohe:PPv2}, which includes spin precession.
These exceptions are GW190413\_052954, GW190426\_152155, and GW190909\_114149.
We then choose subsets of the original broad template bank from the GWTC-2 analysis by comparing against the posteriors of each event,
using the following steps as introduced by~\citet{Li:2019osa}:
We first draw $O(1000)$ of each event's posterior samples with the highest likelihoods
to account for the uncertainty in the event's measured mass and spin parameters.
For each sample we simulate,
using the aligned-spin SEOBNRv4\_ROM waveform model~\citep{Bohe:2016gbl, Purrer:2014fza, Purrer:2015tud},
one signal with the event's original optimal signal-to-noise ratio $\rho_\text{opt}$ as given by Eq. (2) in \citep{Li:2019osa}
and nine extra signals with smaller $\rho_\text{opt}$,
scaled by changing their effective distances $D_\text{eff}$~\citep{Allen:2005fk}.
The reduced template bank for an event is then constructed by searching the simulated data with the original GWTC-2 template bank
(which also consists of SEOBNRv4\_ROM waveforms) and keeping those templates that recover any of the simulated signals with a \far $<$ $1$ in $30$ days.

For \pycbc we select a single template for each search, choosing the maximum-posterior redshifted masses and aligned-spin components $\{(1+z)m_1, (1+z)m_2, \chi_1, \chi_2\}$
as estimated from a four-dimensional Gaussian \kde fit to the posterior samples from \citet{gwosc:gwtc2} for these parameters.
Where available, we use aligned-spin posterior samples. In the case of GW190412 and GW190814, we use samples generated using the SEOBNRv4\_ROM waveform; for GW190426\_152155 we use a mixture of samples generated using the SEOBNRv4\_ROM\_NRTidalv2\_NSBH and IMRPhenomNSBH waveforms; and for GW190425 we use samples generated using the IMRPhenomD\_NRTidal, TEOBResumS, and SEOBNRv4T\_surrogate waveforms.
If aligned-spin posteriors are not available in the \citet{gwosc:gwtc2} data release, we use precessing posterior samples and marginalise over the transverse-spin components before applying the KDE. This produces an aligned-spin template with high matches at the peak of the posterior.
In the case of GW190521, we use samples generated using the \IMRPHM~\citep{Khan:2019kot}, \NRSur~\citep{Varma:2019csw} and SEOBNRv4PHM~\citep{Ossokine:2020kjp} waveforms.
For all other events, we use samples generated using the SEOBNRv4P and \IMRP waveforms.

These choices of waveforms and posterior samples are not necessarily optimal,
but they are valid for this analysis in the sense that
the recovery of similar waveforms with parameters close to the best-fit ones for the targeted GWTC-2 events has been verified through injection studies.
In addition, in the actual searches, the targeted banks constructed in this way
successfully recovered the corresponding GWTC-2 events in all \gstlal searches,
while for \pycbc triggers within 0.1\,s of the target events were excluded from the final trigger list,
but in all cases where the original events were observed with two or more detector,
a coincident trigger was also recovered in the targeted search.
In future work, revisiting the choice of posterior samples used to construct template banks may further improve sub-threshold searches' effectiveness.

\section{Injection study for microlensing analysis}
\label{app:injection_study_microlensing}
\begin{figure}
\centering
\includegraphics
  [width=0.5\hsize]
  {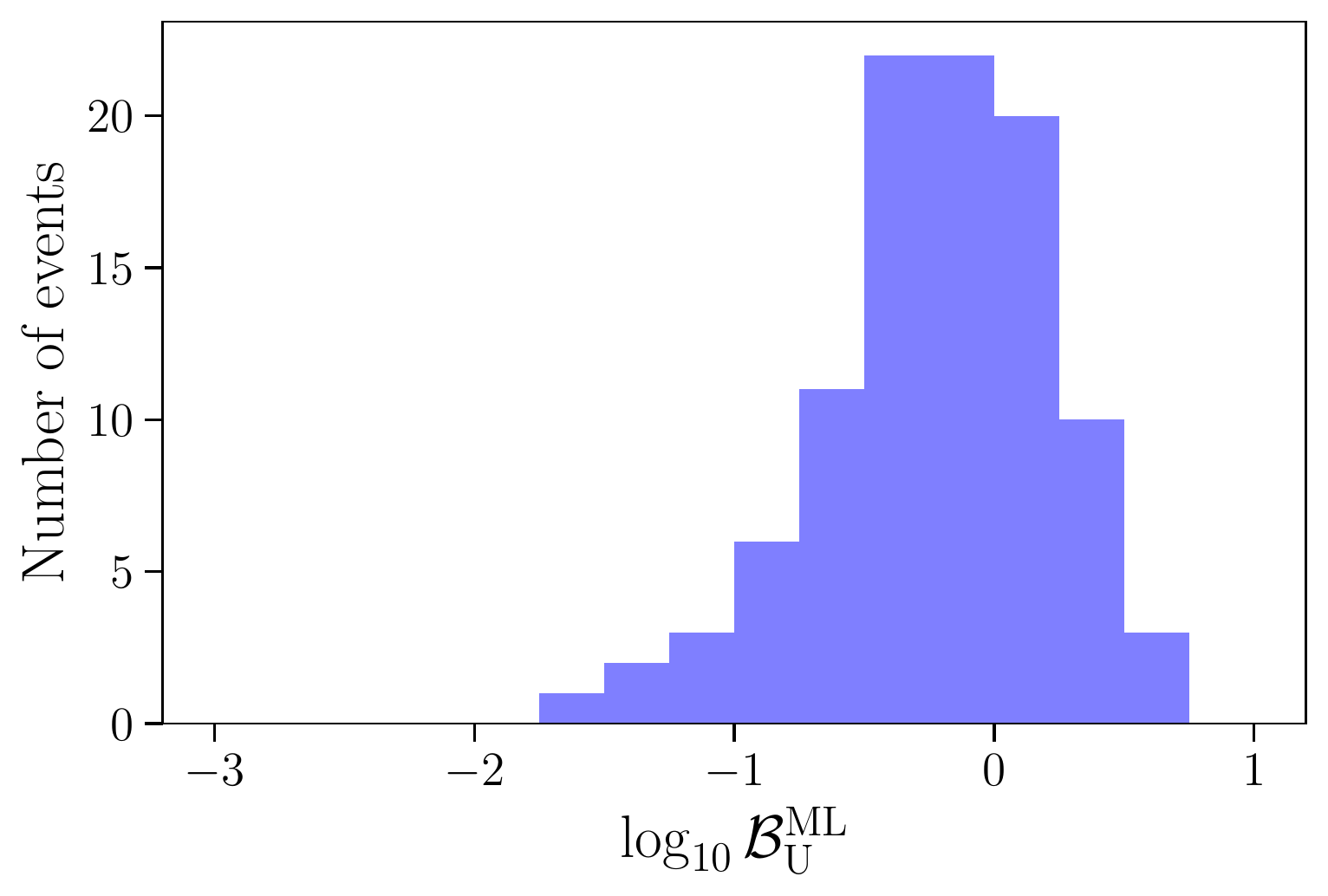}
\caption{
Distribution of microlensing Bayes factors $\log_{10}\BMLU$ for unlensed simulated signals,
recovered using a lensed template.
\label{fig:ijstudy_blu_unlensed}
}
\end{figure}

A high Bayes factor $\BMLU$ itself is not conclusive evidence of microlensing in an observed event.
We have performed an injection study to explore the impact of statistical fluctuations on the Bayes factor obtained from unlensed signals.
We generate unlensed injections by randomly drawing from the parameter space of precessing \bbh systems. 
Simulated Gaussian noise is used considering nominal O3 sensitivity~\citep{Aasi:2013wya}, and we use the \IMRP waveform model~\citep{Hannam:2013oca,Bohe:PPv2} for all simulated injections.
The statistical fluctuations of $\log_{10}\BMLU$ for 100 unlensed injections recovered using lensed templates
can been seen in Fig.~\ref{fig:ijstudy_blu_unlensed}
which shows that the typical values found are \mbox{$\log_{10}\BMLU < 0.75$}.

\bibliography{o3a_lensing}

\clearpage

\iftoggle{endauthorlist}{
 \let\author\myauthor
 \let\affiliation\myaffiliation
 \let\maketitle\mymaketitle
  
 \pacs{}
 \maketitle
}

\end{document}